\documentclass[12pt, a4paper]{article}

\usepackage[left=2.00cm, right=2.00cm, top=3.00cm, bottom=2.00cm]{geometry}
\usepackage{amsmath, amssymb, graphicx}
\usepackage{amsfonts}
\usepackage{makeidx}
\usepackage[english]{babel}
\usepackage{jheppub}
\usepackage[normalem]{ulem}
\usepackage{color}

\newcommand{\be}{\begin{equation}}
\newcommand{\ee}{\end{equation}}
\newcommand{\bea}{\begin{eqnarray}}
\newcommand{\eea}{\end{eqnarray}}

\newcommand{\nn}{\nonumber}

\newcommand{\fS}{\mathfrak{S}}

\title{Holographic control of information \\
	and dynamical topology change \\
	for composite open quantum systems}
\author{Irina Aref'eva,}
\author{Oleg Inozemcev,}
\author{Igor Volovich}
\affiliation{Steklov Mathematical Institute, Russian Academy of Sciences, Gubkin str. 8, 119991
Moscow, Russia}

\emailAdd{arefeva@mi.ras.ru}
\emailAdd{volovich@mi.ras.ru}
\emailAdd{inozemcev@mi.ras.ru}

\abstract{We investigate how the compositeness of a quantum system influences the characteristic time of equilibration. We study the dynamics of open composite quantum systems strongly coupled to the environment after a quantum perturbation accompanied by non-equilibrium heating. We use a holographic description of the evolution of entanglement entropy. The non-smooth character of the evolution with holographic entanglement is
a general feature of composite systems, which demonstrate a dynamical change of topology in the bulk
space and a jump-like velocity change of entanglement entropy propagation. Moreover, the number of jumps
depends on the system configuration and especially on the number of composite parts. The evolution of
the mutual information of two composite systems inherits these jumps. We present a detailed study of
the mutual information for two subsystems with one of them being bipartite. We have found 5 qualitatively
different types of behavior of the mutual information dynamics and indicated the corresponding ranges of
system parameters.}

\begin{document}
\maketitle

\newpage
\section{Introduction}

In recent years, there has been a growing interest in studying quantum entanglement entropy and
quantum mutual information for various open quantum systems under conditions of environmental change
and with control of it (see \cite{OhyaVol} and the references therein).

In \cite{AreVol-Photo}, the holographic approach was applied to photosynthesis, which is an important example of non-trivial quantum phenomena relevant for life and is studied in the emerging field of quantum biology \cite{OhyaVol,QEB}. Light-harvesting complexes of photosynthetic organisms are many-body quantum systems in which quantum coherence has recently been experimentally shown to survive for relatively long time scales at a physiological
temperature despite the decoherence effects of the environment.

There are successful applications of the holographic AdS/CFT correspondence to high energy and condensed matter physics \cite{Aharony:1999ti, IA, IA15}. In \cite{AreVol-Photo}, the holographic approach was used to evaluate the time dependence of entanglement entropy and quantum mutual information in the Fenna--Matthews--Olson complex during
the transfer of an excitation from a chlorosome antenna to a reaction center. It was demonstrated that the evolution of the quantum mutual information simulating the behavior of solutions of the GKSL master equation in some cases can be obtained using dual gravity describing black hole formation in the AdS space–time with the Vaidya metric (hereafter called AdS-Vaidya spacetime). Estimates of the wake-up and scrambling times for various decompositions of the Fenna--Matthews--Olson complex were obtained.

Many photosynthetic light-harvesting complexes are conventionally modeled by a general three-part structure comprising an antenna, a transfer network, and a reaction center. The antenna captures sunlight photons and excites pigment electrons from their ground state. The excited electrons, which combine with holes to form excitons, travel from the antenna through the intermediate transfer complex to the reaction center, where they participate in chemical reactions.

Here, based on the results in \cite{AreVol-Photo} and \cite{DAIA1, DAIA2}, we study whether the compositeness of the system increases the delocalization during equilibration. For this, we consider the dynamics of open quantum
systems consisting of separate parts strongly coupled to the environment after a quantum perturbation corresponding to nonequilibrium heating (also see \cite{KV}--\cite{AHT}). We use a holographic description of the time evolution of entanglement entropy during the nonequilibrium heating.

A nonsmooth character of the evolution is a general feature of the time evolution of the holographic
entanglement of composite systems. These systems exhibit a dynamical topology change in the bulk space
and also a jumplike change of the velocity of the entanglement entropy propagation, and the number of
jumps depends on the system configuration and especially on the number of composite parts. The evolution
of the mutual information of two composite systems inherits these discontinuities.

We present a detailed study of the mutual information for two subsystems, one of which is bipartite.
We show that there exist five qualitatively different types of dynamical behavior of the mutual information
and indicate the corresponding regions of the system parameters.

\vspace{5ex}
\section{Setup}

\subsection{Holographic entanglement entropy in Vaidya-AdS$_3$}
We use a holographic approach to study evolution of an open system after a quantum quench accompanied by non-equilibrium heating process. As the dual model describing the time evolution of the entanglement during such a process, we consider a Vaidya shell collapsing on a black hole \cite{AreVol-Photo}, \cite{Bala}. The collapse of this shell leads to the formation of a heavier black hole, which corresponds to a temperature increase. The initial thermal state is defined as the horizon position $z_H$, and the final state as the horizon $z_h$. For simplicity, we consider the three-dimensional case. The corresponding Vaidya metric defining the dual gravitational background consists of two parts and is given by

\begin{eqnarray}
\label{Vm1}
v<0:&\,\,\,&
ds^2=\frac{1}{z^2}\left( -f_{H}(z)d\tau^2+ \frac{dz^2}{f_{H}(z)}+dx^2\right),\,\,\,\,\tau=v+z_H \text{arctanh} \frac{z}{z_H},\\
\label{Vm2}v>0:&\,\,\,&
ds^2=\frac{1}{z^2}\left( -f_{h}(z)d\tau^2+ \frac{dz^2}{f_{h}(z)}+dx^2\right),\,\,\,\,\,\,\tau=v+z_h \text{arctanh} \frac{z}{z_h},
\end{eqnarray}
where the functions $f_{H}$ and $f_{h}$ are defined as
\begin{eqnarray} \label{BHpm}
f_H = 1-\left(\frac{z}{z_{H}}\right)^2, \,\,\,\,
f_{h} = 1-\left(\frac{z}{z_{h}}\right)^2, \,\,\,\, 0<z_h<z_{H}.
\end{eqnarray}
and are glued along the position of the shell $v=0$. This metric can also be written as
\begin{align}
\label{Vm1a}
v<0: ~~~ ds^2 = \frac{1}{z^2}\left[ -\left(1-\frac{z^2}{z_H^2}\right)dv^2 - 2\,dv\,dz + dx^2 \right], \\
\label{Vm2a}
v>0: ~~~ ds^2 = \frac{1}{z^2}\left[ -\left(1-\frac{z^2}{z_h^2}\right)dv^2 - 2\,dv\,dz + dx^2 \right]. 
\end{align}

The initial and finite temperatures are
\begin{equation}
T_i=\frac{1}{2\pi z_H},~~~~ T_f=\frac{1}{2\pi z_h}.
\end{equation}	

To define the holographic entanglement entropy \cite{RTderiv,Hubeny:2007xt} corresponding to the simplest system, i.e., one segment, we must find a geodesic in metric \eqref{Vm1a}, \eqref{Vm2a} anchored on this segment at a given instant.
The action for the geodesic connecting two points $(t,-\ell/2)$ and $(t,\ell/2)$ on the boundary is given by
\begin{equation} \label{action1}
S(\ell,t)=\int\limits_{-\ell/2}^{+\ell/2}\frac{1}{z}\,\sqrt{1-2v'z'-f(z,v)v^{\prime 2}}\,dx,
\end{equation}
where $t\geq0$, $\ell>0$, $z=z(x)$, $v=v(x)$ and
\begin{equation} \label{BC}
z(\pm \ell/2)=0,\,\,\,\,v(\pm \ell/2)= t.
\end{equation}
To realize these symmetric boundary conditions, we also impose the conditions $z'(0)=v'(0)=0$.

This problem was solved explicitly for a non-zero initial temperature in \cite{DAIA1,DAIA2} and the answer is given by\footnote{The entanglement entropy \eqref{fS} has been obtained as a length of the geodesic with minimal divergence subtraction.} 
\be 
\label{S-AA}S(\ell,t)= \log  \left( \frac{z_h} {\ell \,\fS_\kappa (\rho,s)}\,\sinh\frac{t}{z_h}\right), \,\,\,\,\,0\leq t\leq \ell,
\ee
where $\fS_\kappa$ is given by
\be\label{fS}
\fS_\kappa (\rho,s)=\frac {c \rho + \Delta} {\Delta}\cdot\sqrt {\frac {\Delta^2 - 
		c^2 \rho^2} {\rho\left (c^2 \rho + 
		2 c\Delta + \rho \right) - \kappa^2}}.\ee
The function $\fS_\kappa (\rho,s)$ depends on the new variables $(\rho,s)$, which are related to the variables $(\ell,t)$. 
The new variables $\rho,s$ describe the geodesic in the bulk space and are expressed in terms of the position of the geodesic top $z_*$ and the point $z_c$ where the geodesic crosses the lightlike shell by the formulas
\be\label{s-rho}
s=\frac{z_c}{z_*}, \;\;\;\; \rho=\frac{z_h}{z_c},\ee
with the restrictions $z_*<z_H$ and $z_c<z_H$ satisfied.

The relations between $\rho,s$ and $\ell,t$ can be written in the explicit form

\bea\label{ttt}
\frac{t}{z_h}&=&{\mbox {arccoth}}\left(\frac{-c \kappa ^2+2 c \rho
	^2+c+2 \Delta  \rho }{2 c \rho +2 \Delta }\right),
\\
\ell&=&\frac{z_h}{2}  \log \frac{c^2 \gamma ^4-4 \Delta 
	\left(c s \left(\kappa ^2-2 \rho ^2+1\right)+\Delta
	+\Delta  \left(\rho ^2-2\right) s^2\right)}{c^2 \gamma
	^4-4 \Delta ^2 (\rho  s-1)^2}+\frac{z_h}{2 \kappa } \log
\frac{(c \kappa +\Delta  s)^2}{\rho ^2 s^2-\kappa
	^2},\nn\\
\label{ell}\eea
where  we use the notations
\bea\label{Smin}
\kappa=\frac{z_h}{z_H}<1,\,\,\,\,
c=\sqrt{1-s^2},\,\,\,\,
\gamma=1-\kappa^2,\,\,\,\,
\Delta=\sqrt{\rho^2-\kappa^2}.\eea 

For $t<0$, the geodesic is entirely in the bulk space of the black hole with temperature $T_i$. The entanglement entropy is independent of $t$ and equal to 
\be
\label{S-AA}S_i= \log \left( \frac{z_H} {\ell}\sinh\frac {\ell} {z_H}\right)
\ee
after the minimal renormalization.

The top of the corresponding geodesic $z_*<z_H$, and $z_*\to z_H$ as $\ell$ increases in correspondence with \cite{Hubeny:2013hz}.  

We are interested in calculating $S=S(\ell,t)$. Hence, we have to find $\rho=\rho(\ell,t)$ and $s=s(\ell,t)$ first.
From the equation \eqref{ttt} it is possible to find $\rho=\rho(s,t)$. Inserting this expression for $\rho$ into \eqref{ell} we obtain $\ell=\ell((s,\rho(s,t))=\ell(s,t)$, from which we get numerically $s=s(\ell,t)$ and $\rho=\rho(s(\ell,t),t)=\rho(\ell,t)$. And finally we can get $S=S(s(\ell,t),\rho(\ell,t))=S(\ell,t)$. 

Let $\ell$ be fixed and the time increases. At very small $t>0$, the geodesic intersects the null shell and for $t \ll z_h$ the point of intersection is close to the boundary, $z_c\ll z_h$.  When $t$ is of the order of $z_h$, the geodesic intersects the shell behind the horizon, i.e. $z_c>z_h$, and finally at some time the geodesic is  entirely in the black hole (with the temperature $T_f$).

\begin{figure}[ht!]
	\centering
	\includegraphics[scale=0.39]{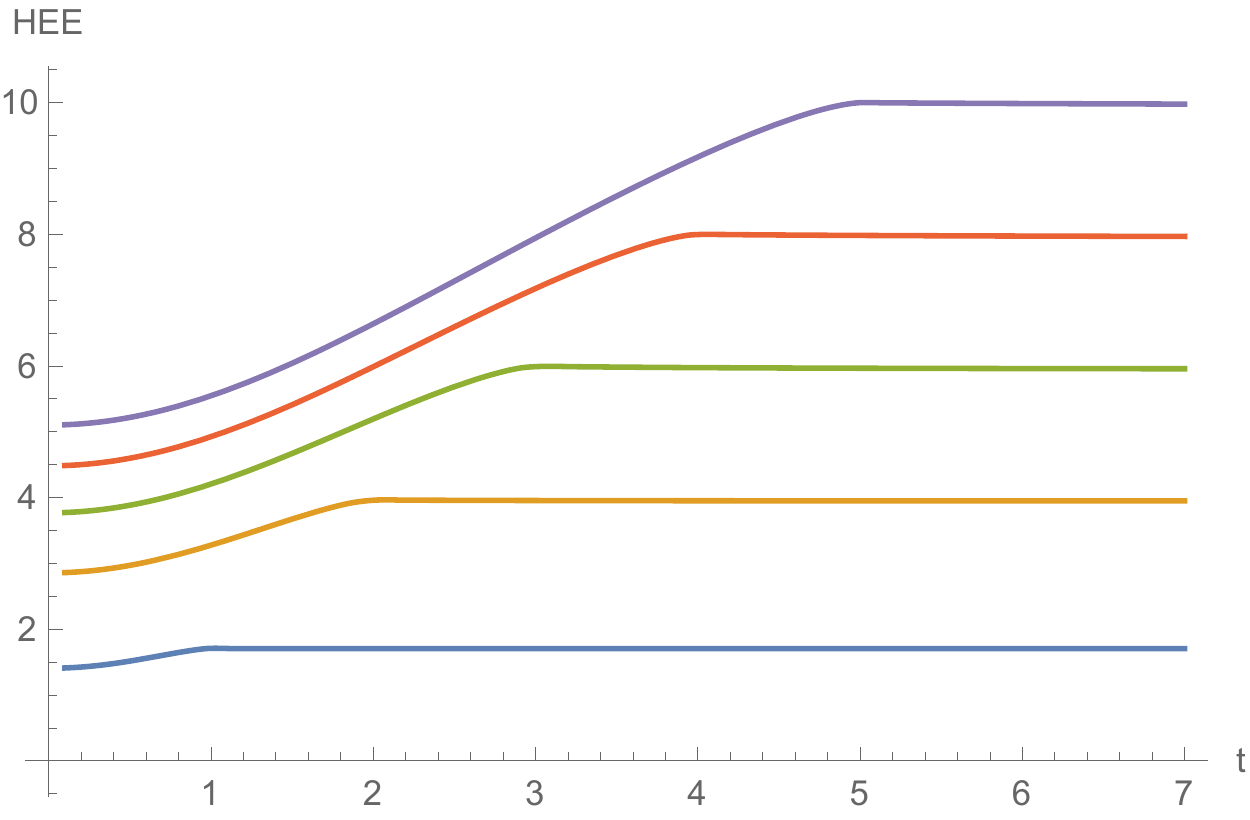}A \hspace{15mm}
	\includegraphics[scale=0.4]{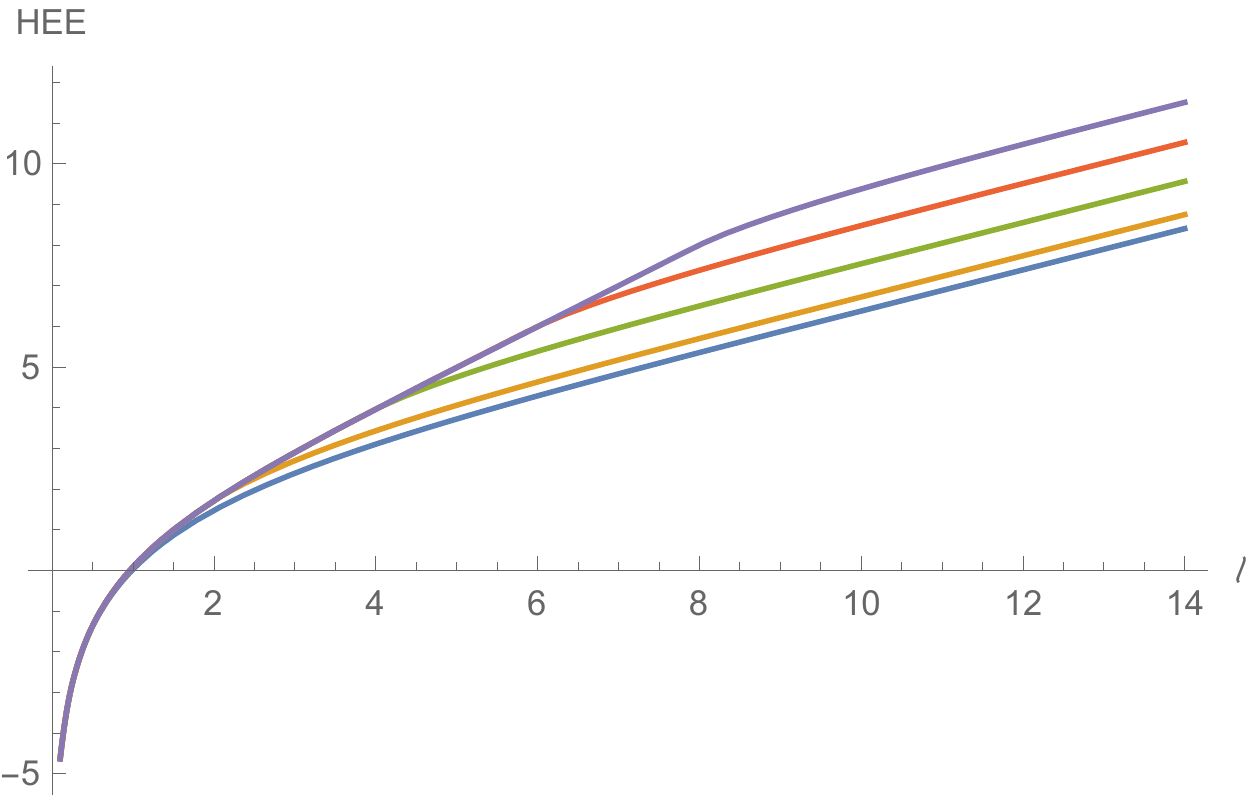}B
	\caption{\textbf{A.} $z_H=4$, $z_h=1$, $\ell$ equals 2, 4, 6, 8, 10 and we vary $t$. \textbf{B.} $z_H=2$, $z_h=1$, $t$ equals 0, 1, 2, 3, 4 and we vary $\ell$.}
	\label{fig:HEE-typ}
\end{figure}

The typical dependence of the holographic entanglement entropy on $t$ for given $\ell$ and on $\ell$ for given $t$ is presented in Fig.\ref{fig:HEE-typ}.
\\

\subsection{Mutual information}

According to the general definition, holographic mutual information (HMI) is
\begin{equation} \label{MI}
I(A;B)=S(A)+S(B)-S(A \cup B),
\end{equation}
where $S(A)$ is the holographic entanglement entropy (HEE) of A, $S(A \cup B)$ is the holographic entanglement entropy for the union of two subsystems.

This definition can be generalized for $m$ subsystems as
\begin{equation}
I(A_1;A_2\cup ...\cup A_m)=S(A_1)+S(A_2\cup ...\cup A_m)-S(A_1\cup A_2\cup ...\cup A_m).
\label{IAm}
\end{equation}

We now consider the evolution of the mutual information of the system consisting of two subsystems, one of them (simple) is a segment $A$ and the second (composite) comprises two segments $B\cup C$. In this case, formula \eqref{IAm} takes the form
\begin{equation}
I(A; B\cup C)=S(A)+S(B\cup C)-S(A\cup B\cup C).
\end{equation}

\vspace{5ex}
\section{Holographic Entanglement Entropy of Composite\\ Systems}

\subsection{Phase diagrams for entanglement entropy of composite systems}

We consider the system of 3 segments with the lengths $\ell$, $m$, $n$ and the distances $x$, $y$ between adjacent segments. If we omit time dependence for brevity, then the formula of the holographic entanglement entropy is (minimum of 15 items):
\begin{multline}
S(\ell,x,m,y,n)=\min \Big\{
S(\ell)+S(m)+S(n), 
S(\ell)+S(m+y)+S(y+n), \\
S(\ell)+S(m+y+n)+S(y), 
S(\ell+x)+S(x+m)+S(n), \\
S(\ell+x)+S(x+m+y)+S(y+n), 
S(\ell+x)+S(x+m+y+n)+S(y), \\
S(\ell+x+m)+S(x)+S(n), 
S(\ell+x+m)+S(x+m+y)+S(m+y+n), \\
S(\ell+x+m)+S(x+m+y+n)+S(m+y), 
S(\ell+x+m+y)+S(x)+S(y+n), \\
S(\ell+x+m+y)+S(x+m)+S(m+y+n), 
S(\ell+x+m+y)+S(x+m+y+n)+S(m), \\
S(\ell+x+m+y+n)+S(x)+S(y), 
S(\ell+x+m+y+n)+S(x+m)+S(m+y), \\
S(\ell+x+m+y+n)+S(x+m+y)+S(m) \Big\}
\label{S3s}
\end{multline}

\begin{figure}[ht!]
	\centering
	\includegraphics[scale=0.23]{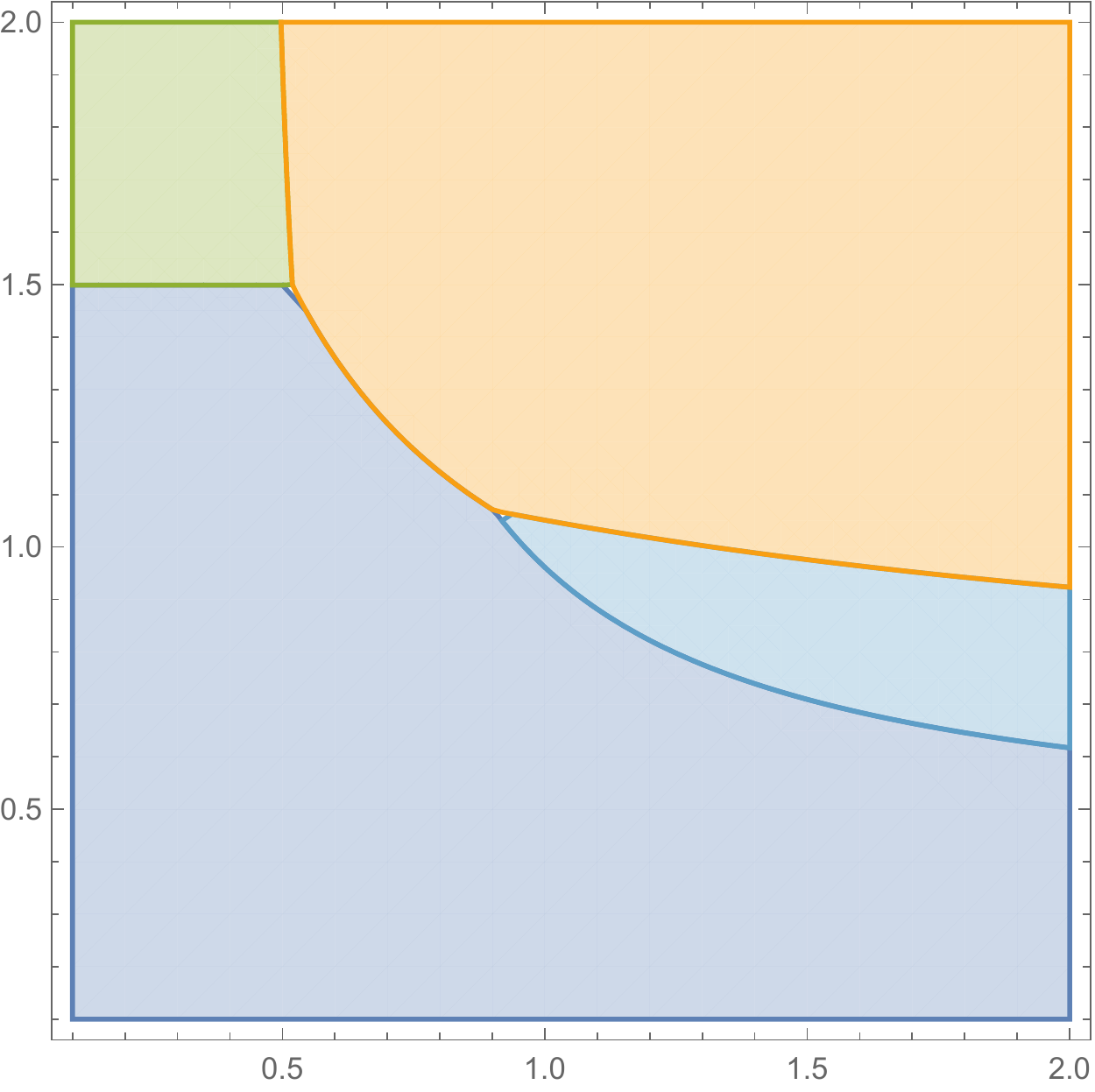}A \;
	\includegraphics[scale=0.23]{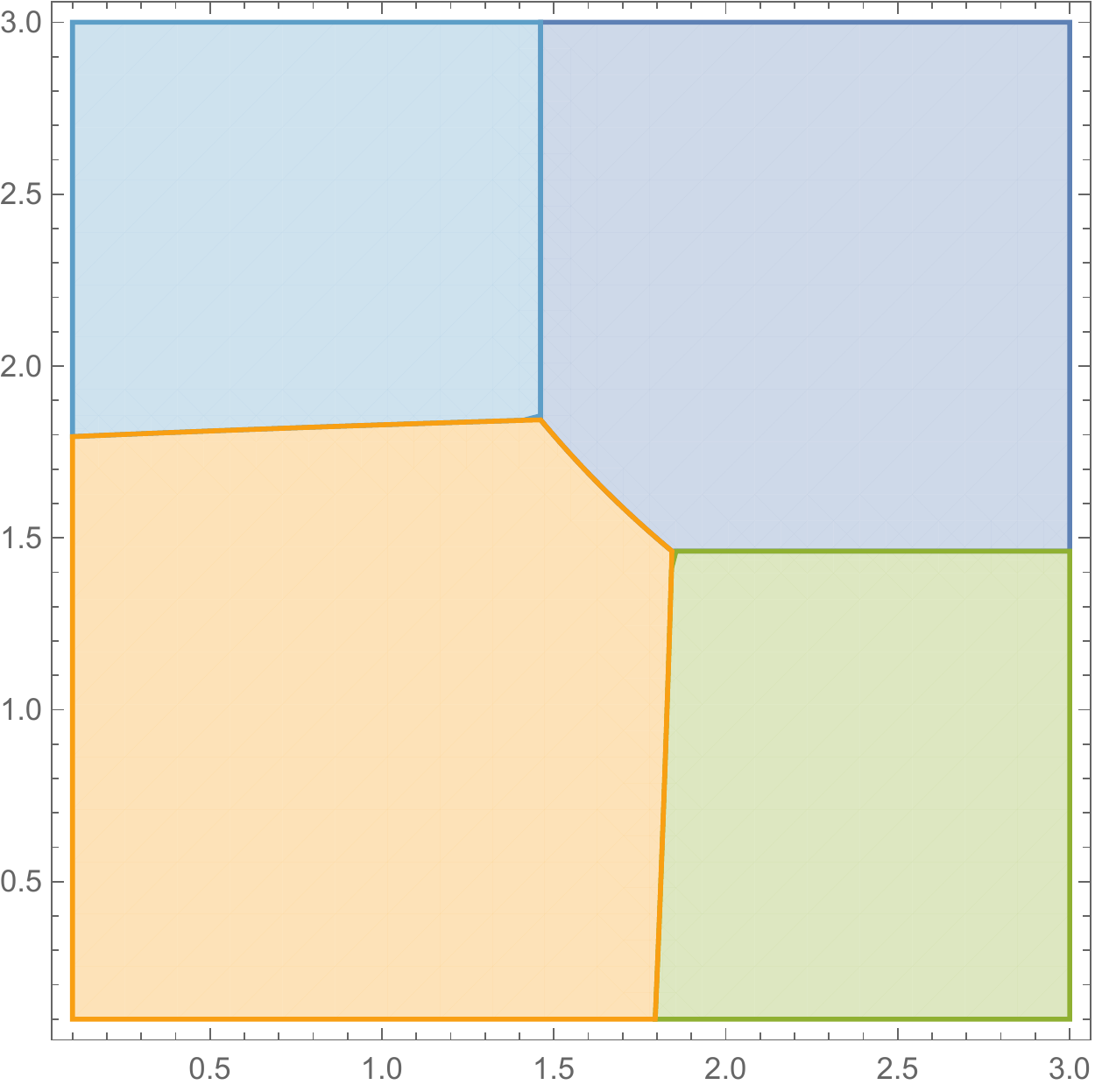}B \;
	\includegraphics[scale=0.23]{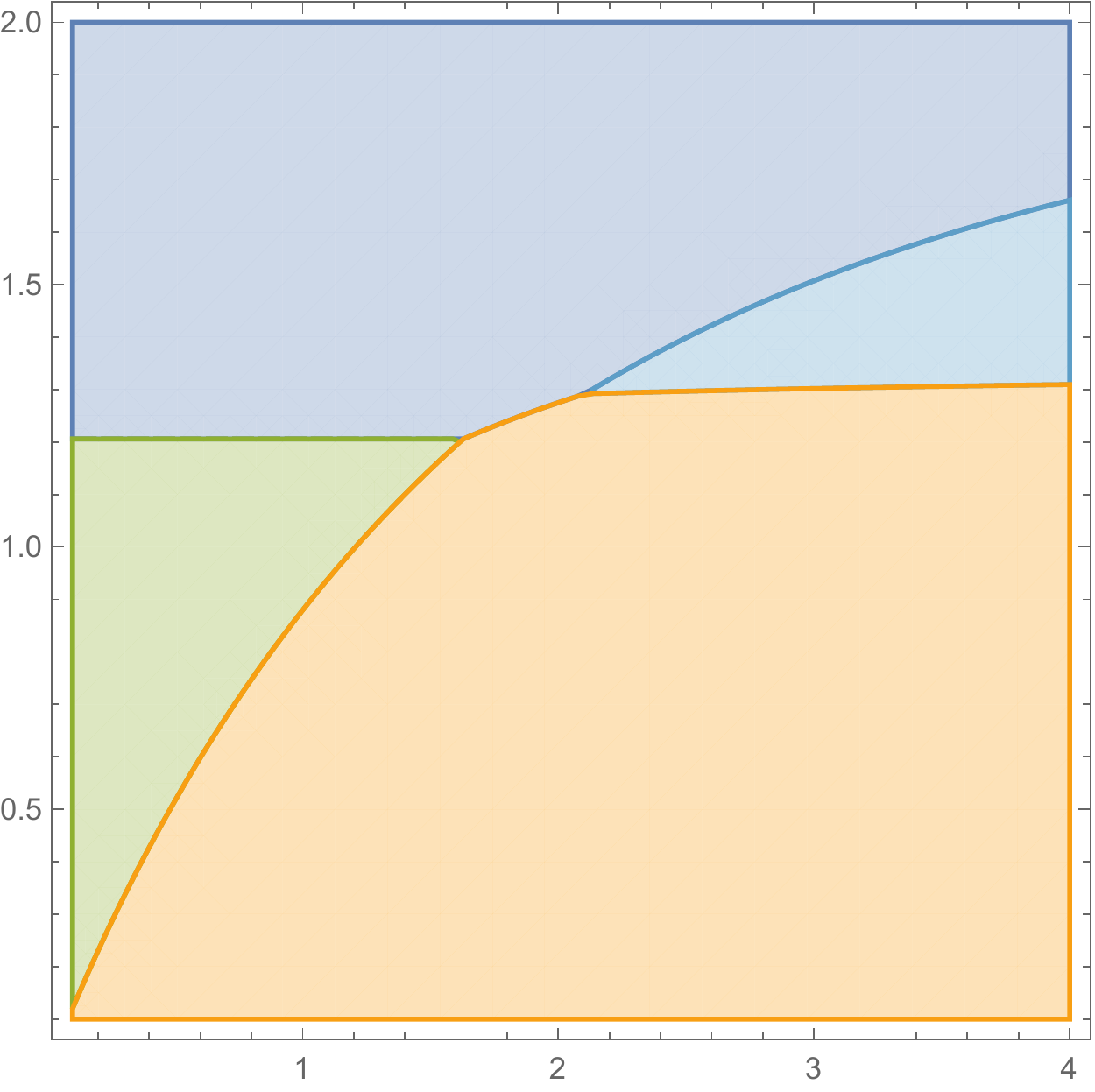}C \;
	\includegraphics[scale=0.23]{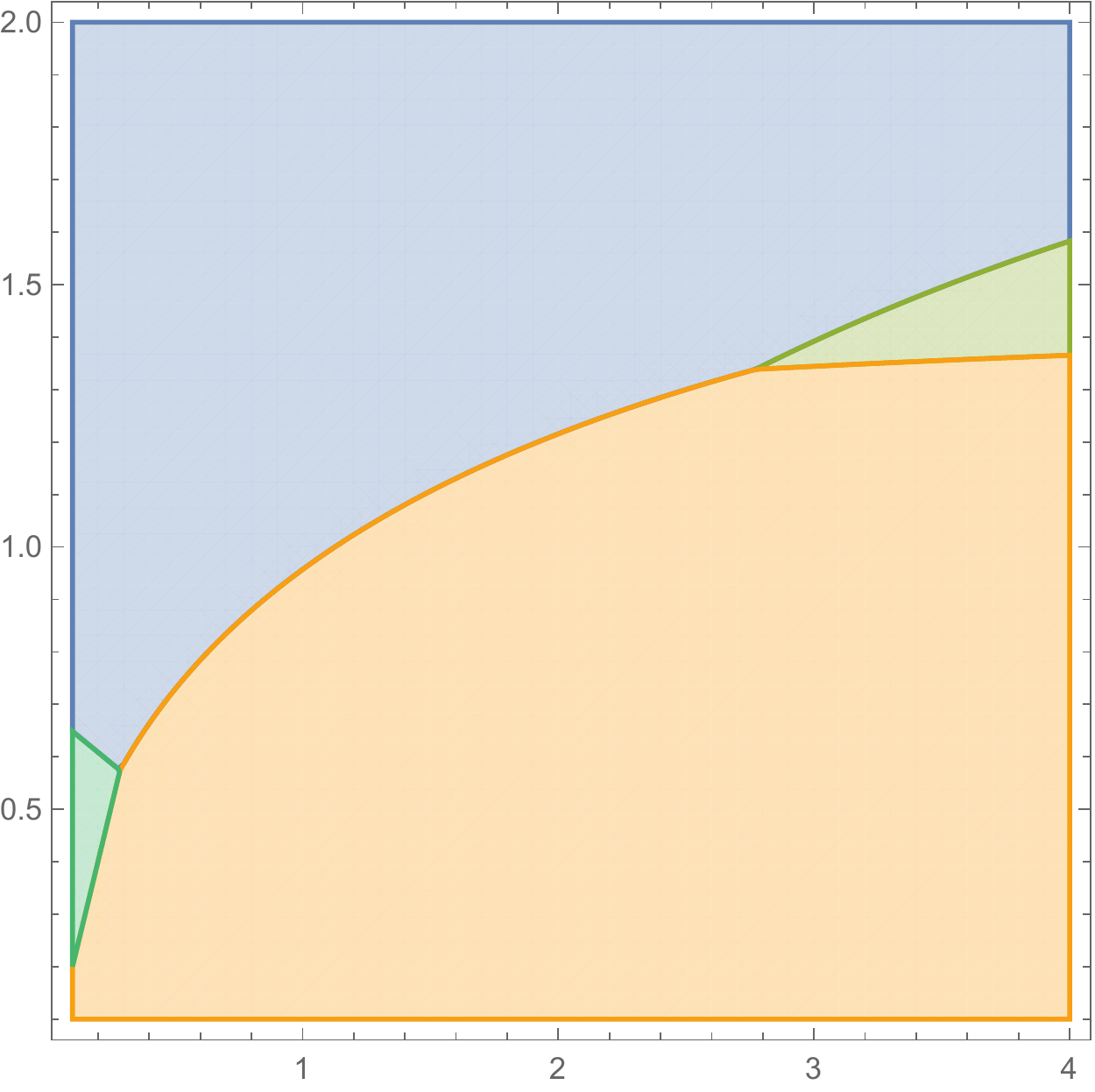}D
	\caption{Phase diagrams for HEE in static background. \textbf{A.} $z_H=3$, $x_1=0.4$, $x_2=0.6$, $l_3=l_2$ and we vary $l_1$, $l_2$. \textbf{B.} $z_H=4$, $l_1=l_2=l_3=4$ and we vary $x_1$, $x_2$. \textbf{C.} $z_H=2.5$, $l_2=5$, $l_3=3$ and we vary $l_1$, $x_2$. \textbf{D.} $z_H=4$, $l_1=l_2=l_3=5$ and we vary $l_2$, $x_2$. On all plots light purple color corresponds to $(A)\|(B)\|(C)$, light green -- $(A)\|(B,C)$, light blue -- $(A,B)\|(C)$, orange -- $(A,B,C)$, darker green (in plot D) -- engulfed configuration.}	
	\label{fig:phase-diag-st}
\end{figure}

Among all configurations in this formula, there are so-called "crossing configurations" (see  \cite{AreVol-Photo}).
Let's check whether crossing configurations contribute to the holographic entanglement entropy in static background. For this purpose, we plot the phase diagrams (Fig.\ref{fig:phase-diag-st}) showing which configurations from \eqref{S3s} contribute to the holographic entanglement entropy. Each region of the solid color corresponds to one configuration of \eqref{S3s}.

\begin{figure}[ht!]
	\centering
	\includegraphics[scale=0.6]{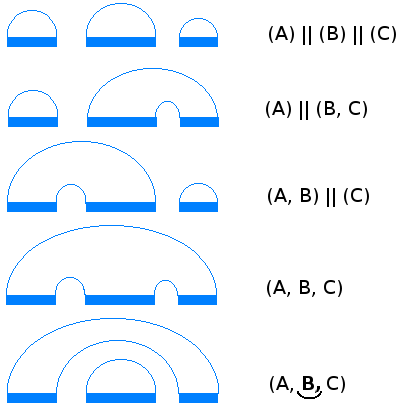}
	\caption{Five contributing configurations of the 3-segment system. $(A)$ is a segment connected to itself, so $(A)\|(B)\|(C)$ is disjoint configuration. $(A, B)$ means that segments $A$ and $B$ connected to each other without intersections. The last one is so called "engulfed" configuration.}
	\label{fig:5-config}
\end{figure}

From Fig.\ref{fig:phase-diag-st} we can see that only the configurations without cross-section contribute to the holographic entanglement entropy. These diagrams represented in Fig.\ref{fig:5-config}. This observation confirms the general statement that crossing configurations do not contribute \cite{Alishahiha:2014jxa,Ben-Ami:2014gsa}.

Without crossing configurations, formula \eqref{S3s} is (minimum of 5 items)
\begin{multline}
S_{no\:cr}(\ell,x,m,y,n)=\min \Big\{
S(\ell)+S(m)+S(n),~ 
S(\ell)+S(m+y+n)+S(y), \\
S(\ell+x+m)+S(x)+S(n),~ 
S(\ell+x+m+y+n)+S(x)+S(y), \\
S(\ell+x+m+y+n)+S(x+m+y)+S(m) \Big\}
\end{multline}

\begin{figure}[ht!]
	\centering
	\includegraphics[scale=0.23]{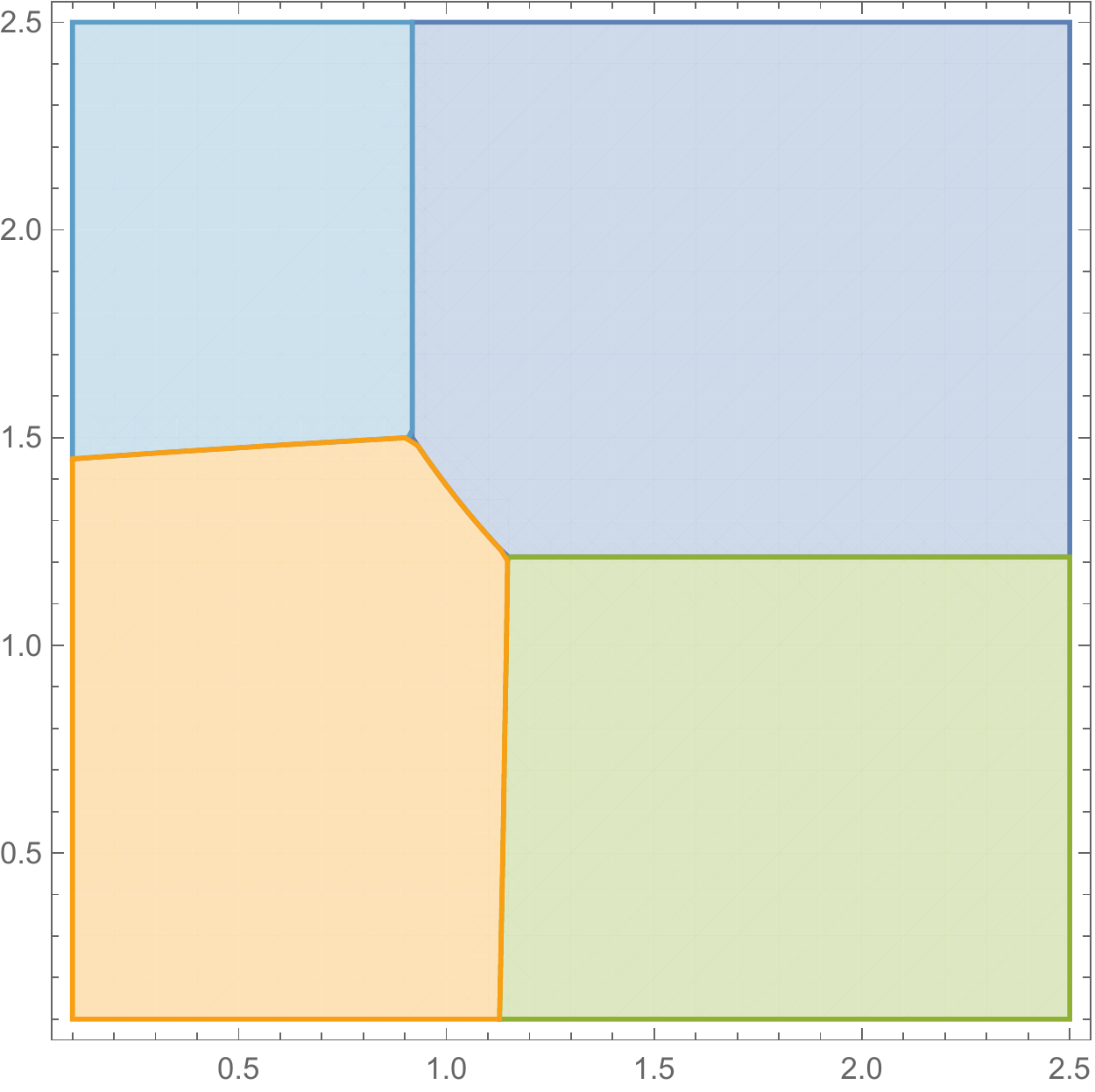} \;\;\;\;
	\includegraphics[scale=0.23]{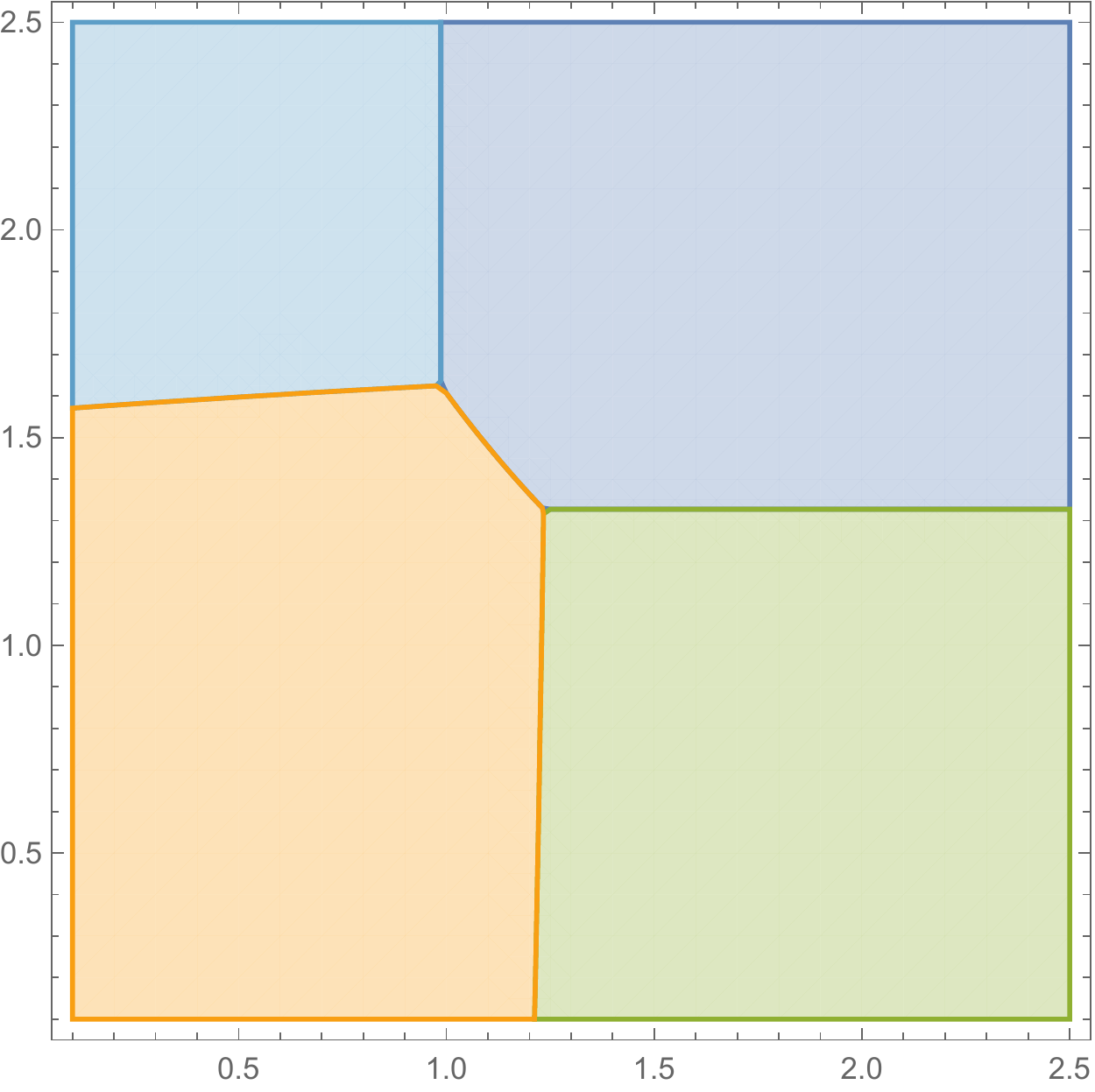} \;\;\;\;
	\includegraphics[scale=0.23]{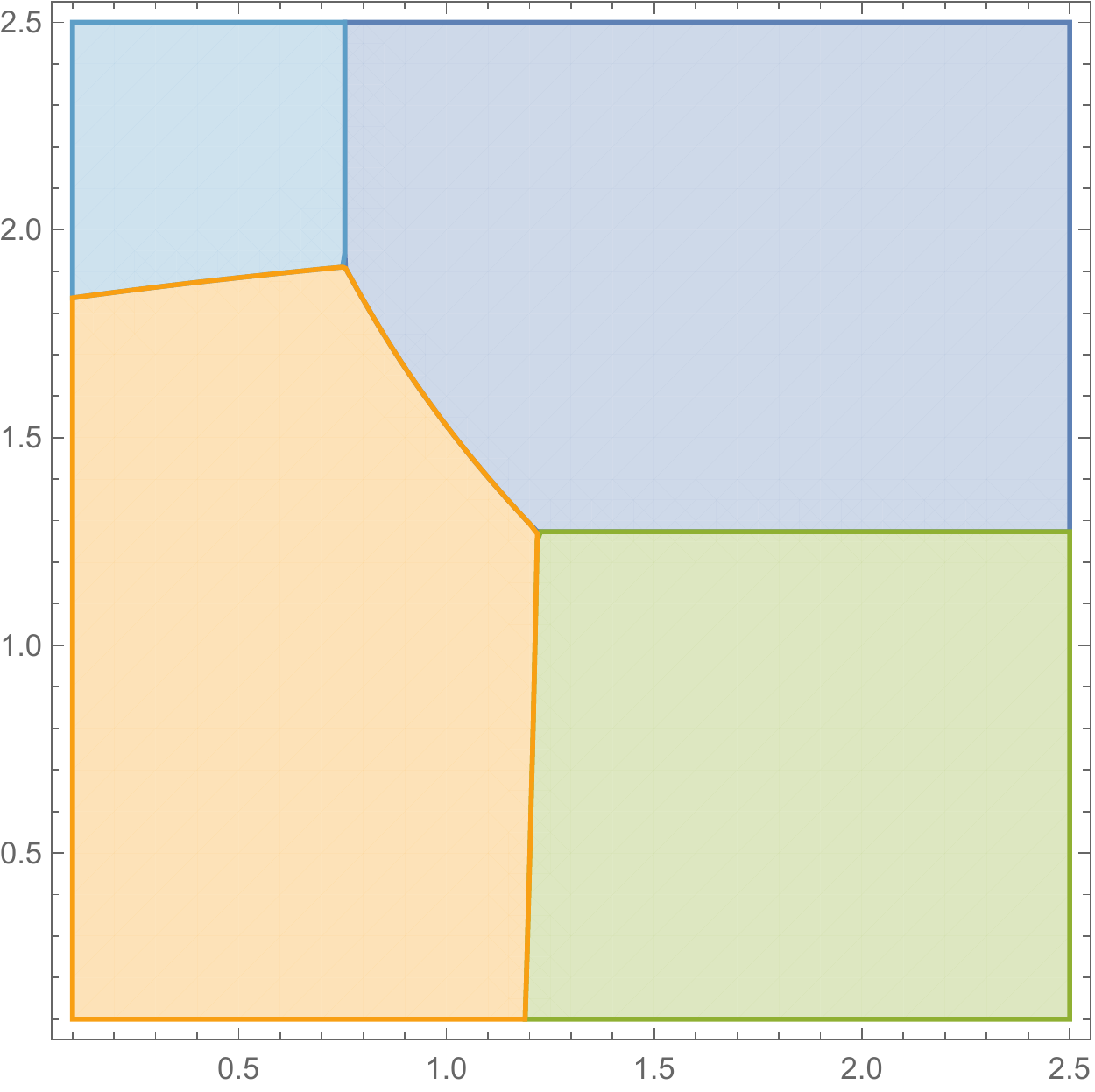} \\ \bigskip
	\includegraphics[scale=0.23]{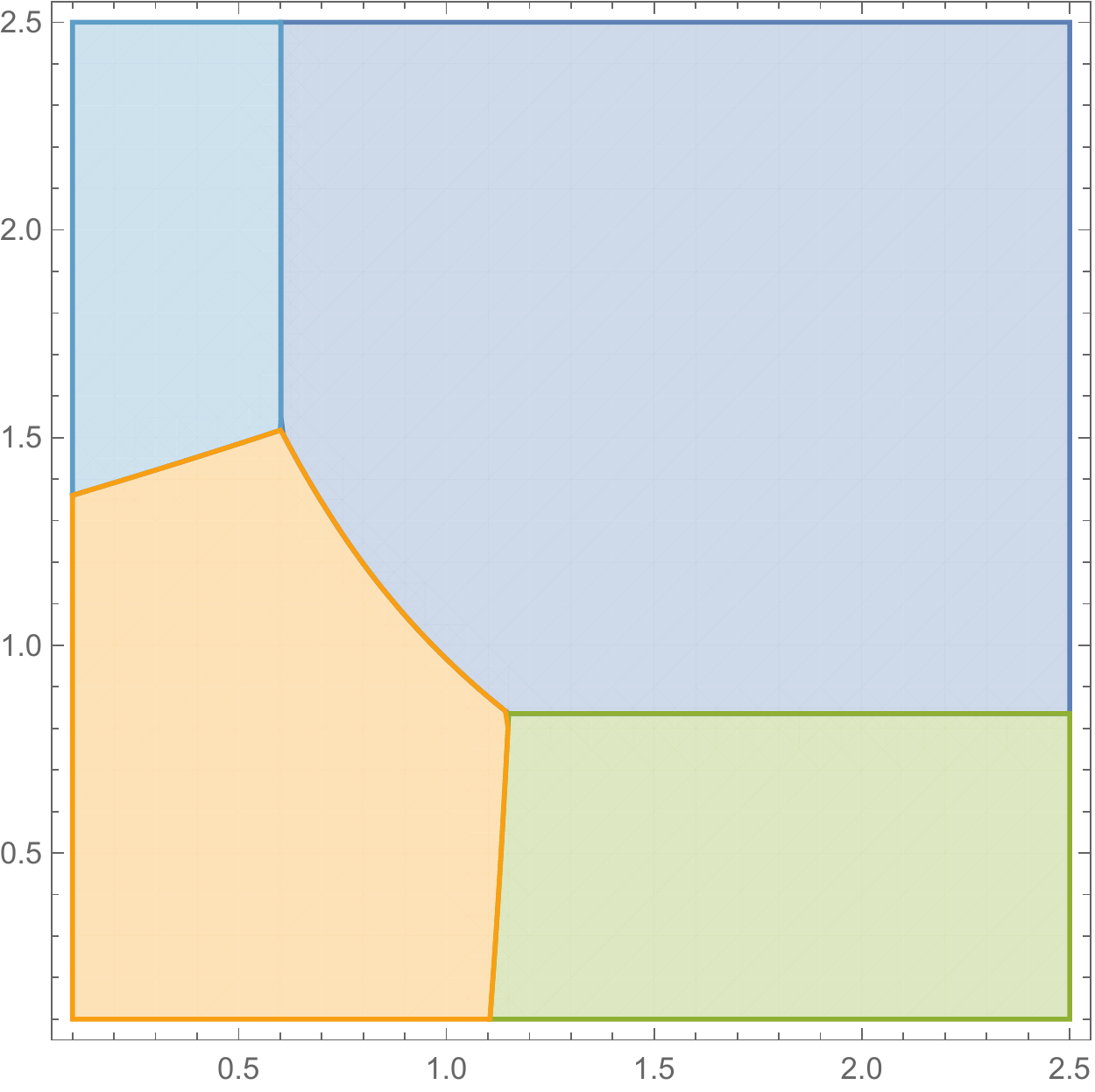} \;\;\;\;
	\includegraphics[scale=0.23]{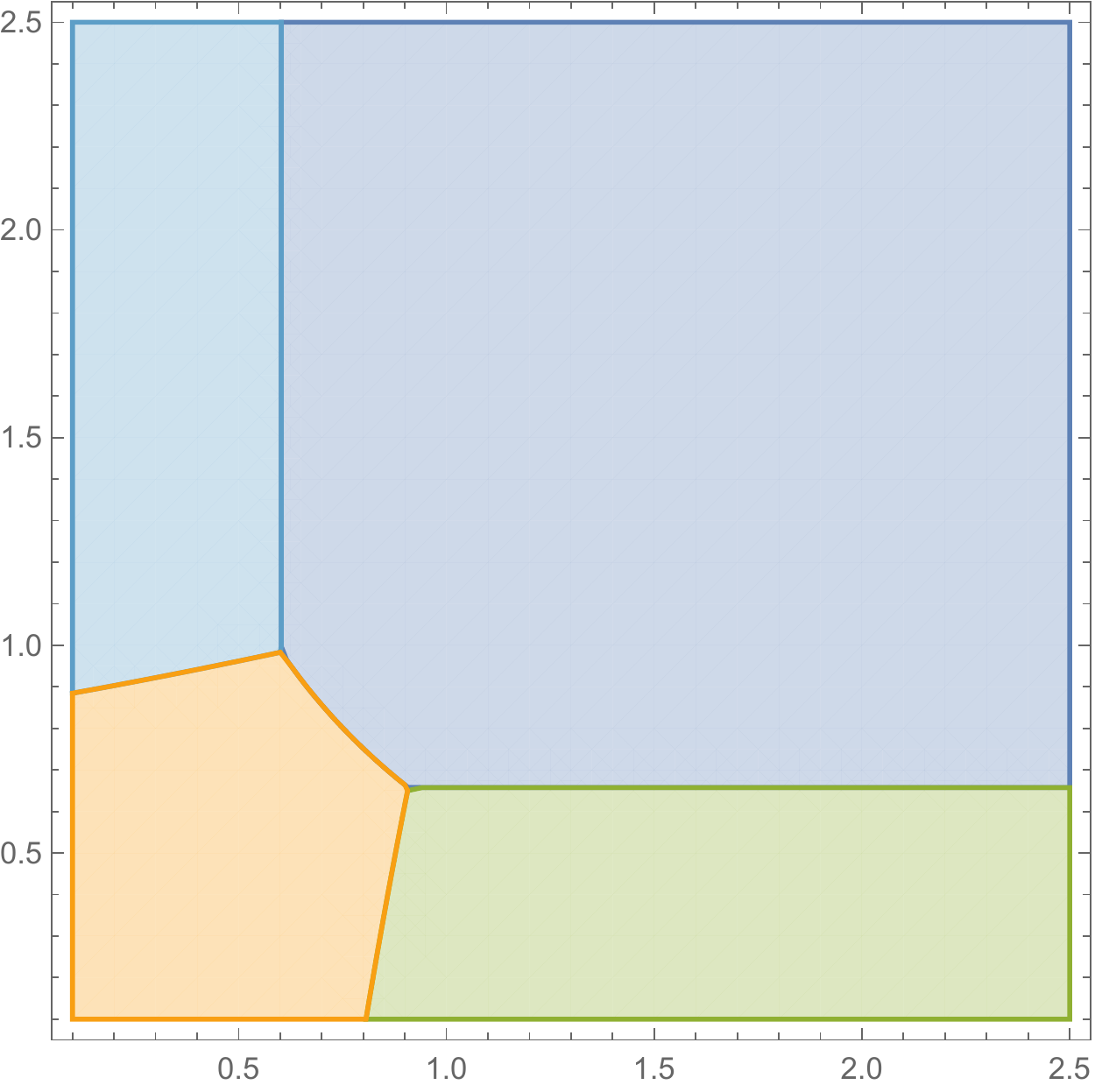} \;\;\;\;
	\includegraphics[scale=0.23]{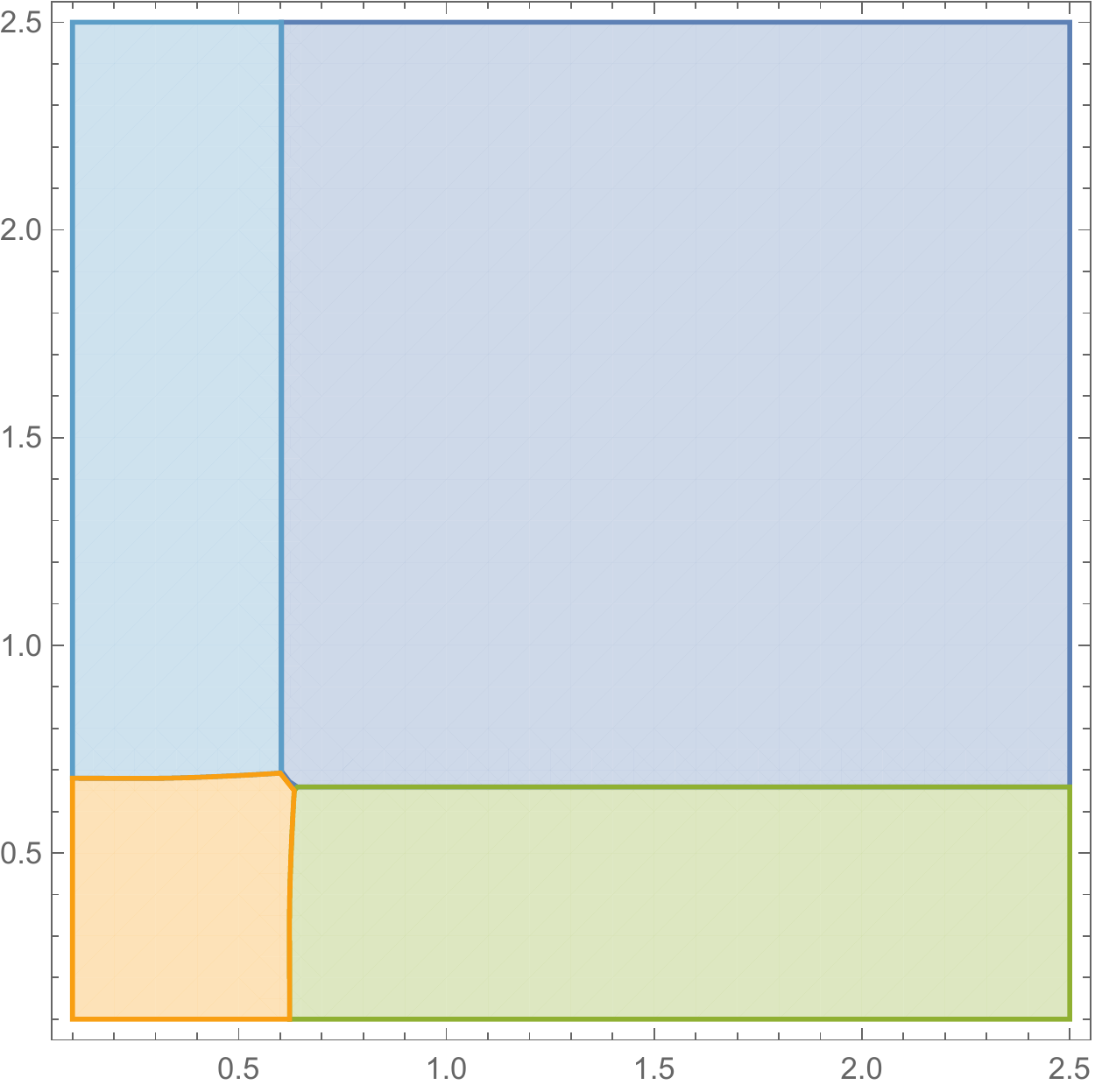}
	\caption{Phase diagrams for HEE in Vaidya-AdS. $z_H=3,~z_h=1,~\ell=2,~m=3,~n=4,~x,y \in (0, 2.5]$, $t=0, 1, 2, 3, 4, 5$. Light purple color corresponds to $(A)\|(B)\|(C)$, light green -- $(A)\|(B,C)$, light blue -- $(A,B)\|(C)$, orange -- $(A,B,C)$.}
	\label{fig:ph-diag-xy}
\end{figure}

\begin{figure}[ht!]
	\centering
	\includegraphics[scale=0.23]{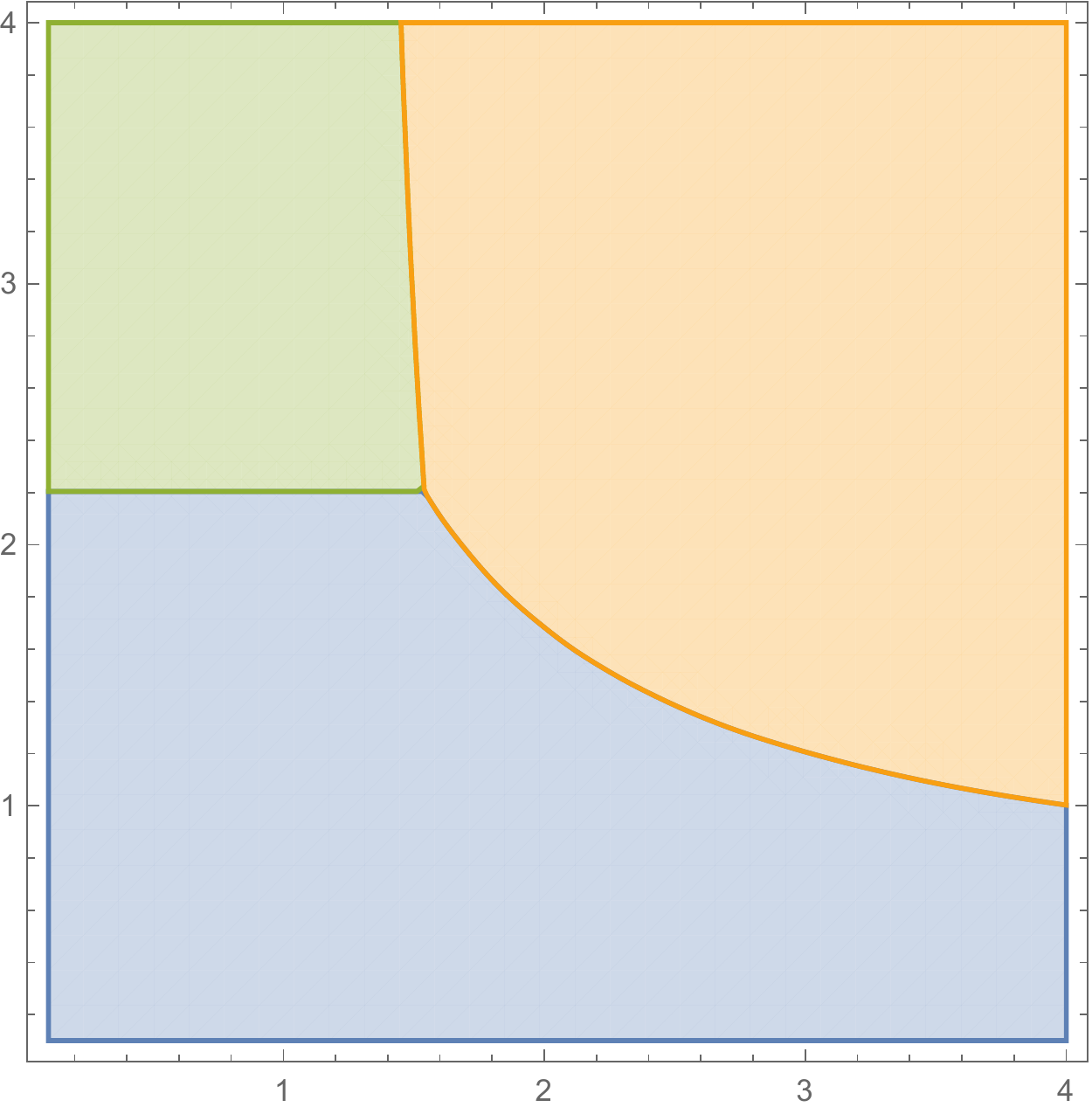} \;\;\;\;
	\includegraphics[scale=0.23]{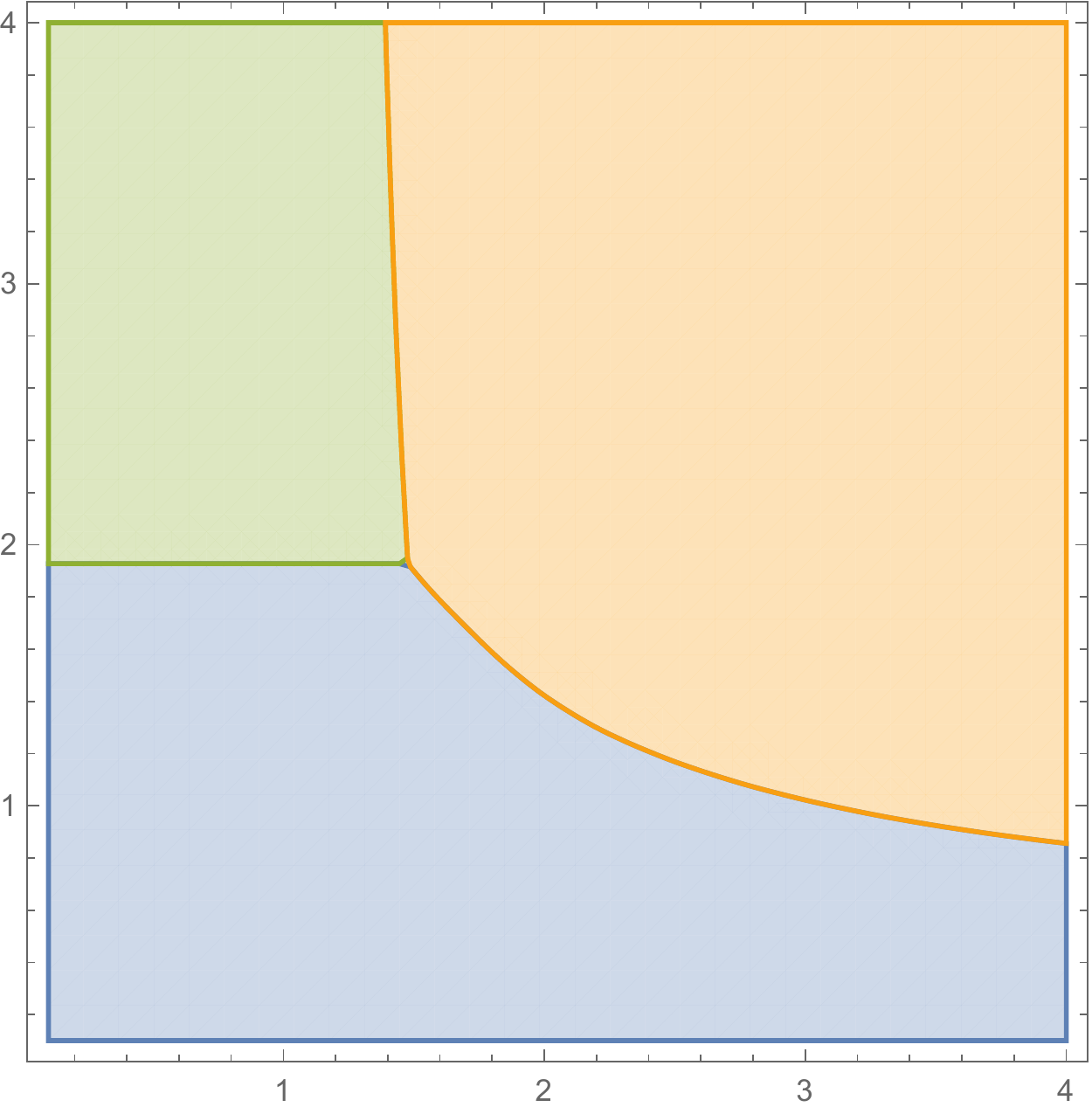} \;\;\;\;
	\includegraphics[scale=0.23]{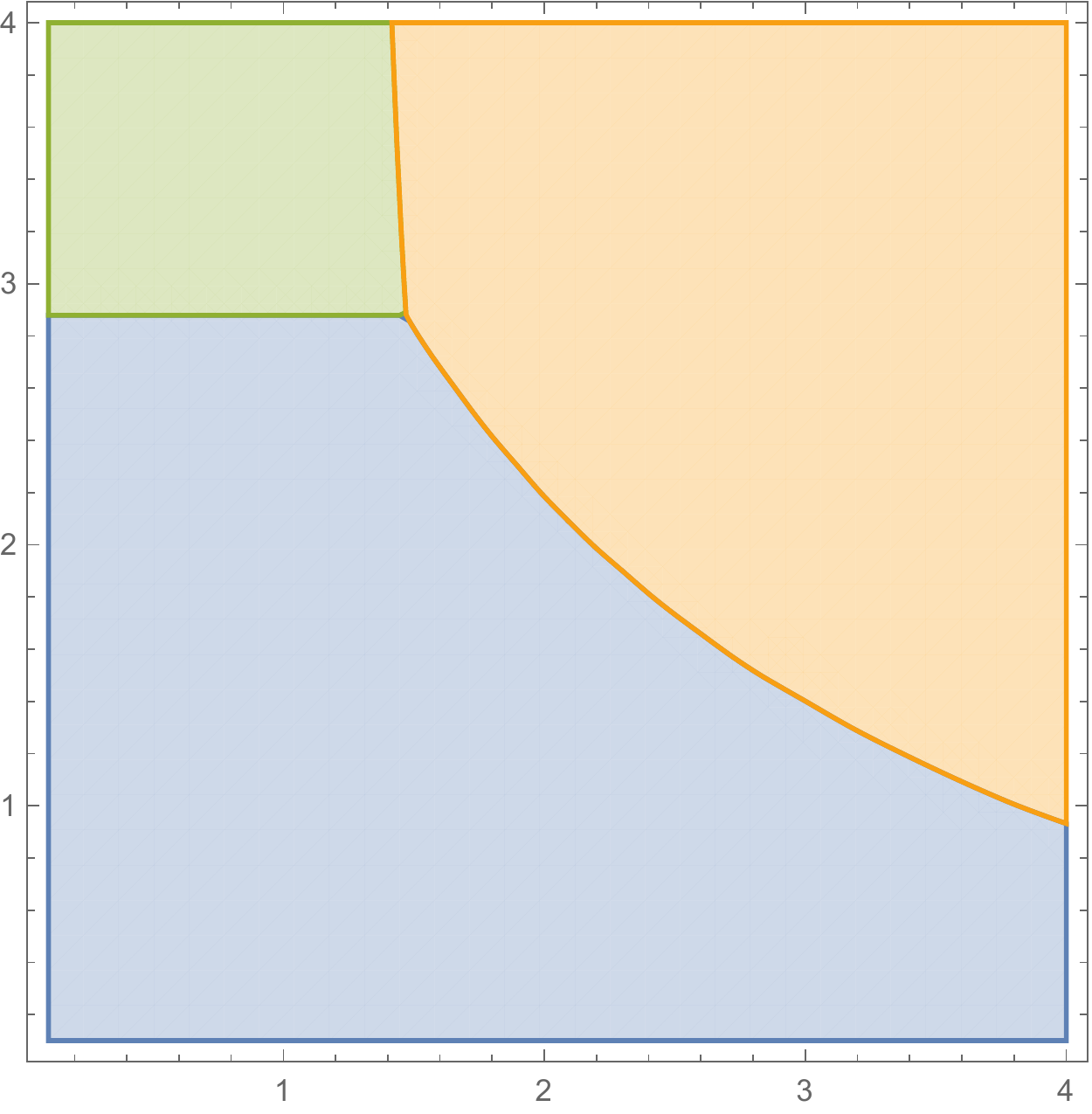} \\ \bigskip
	\includegraphics[scale=0.23]{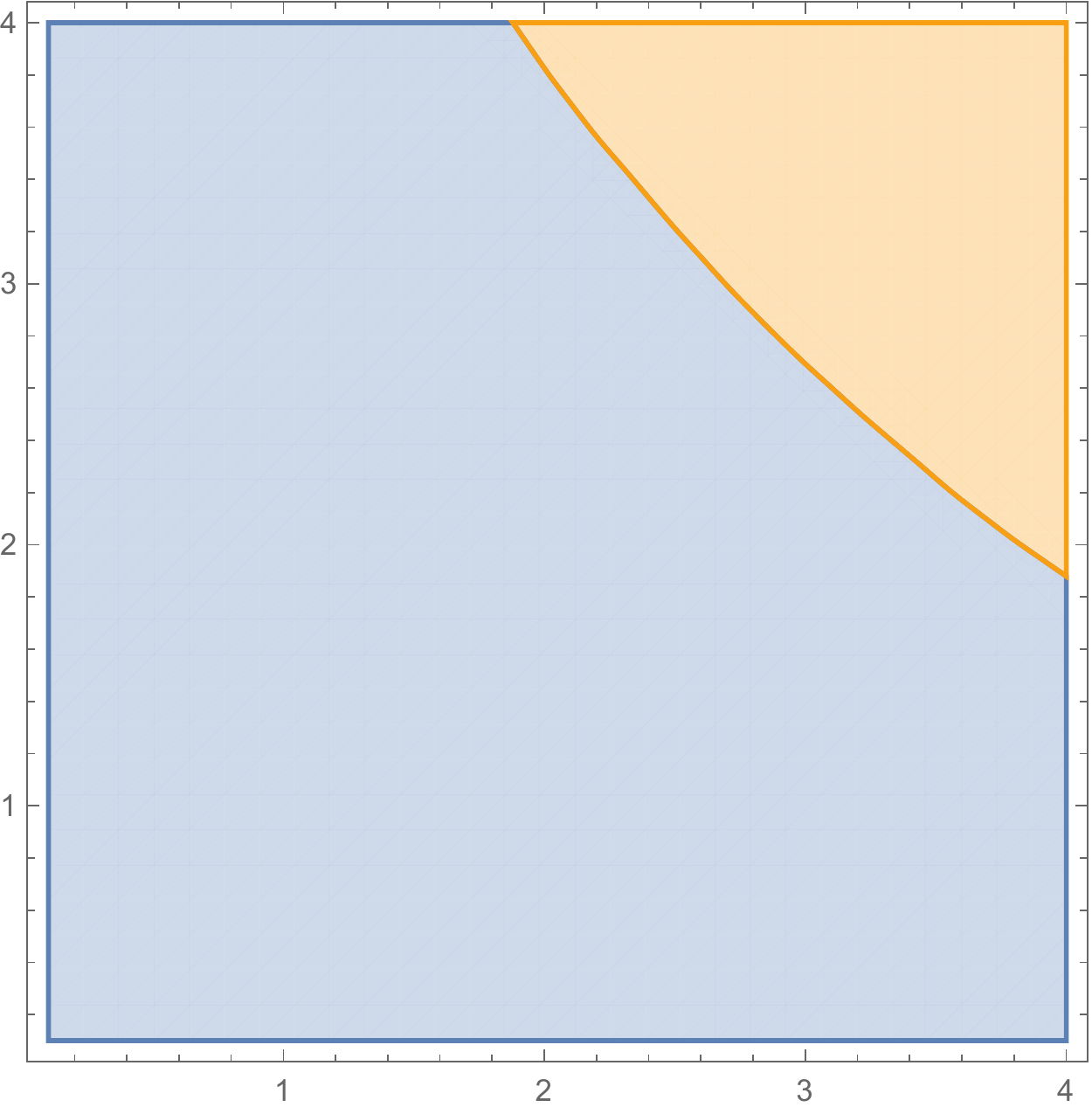} \;\;\;\;
	\includegraphics[scale=0.23]{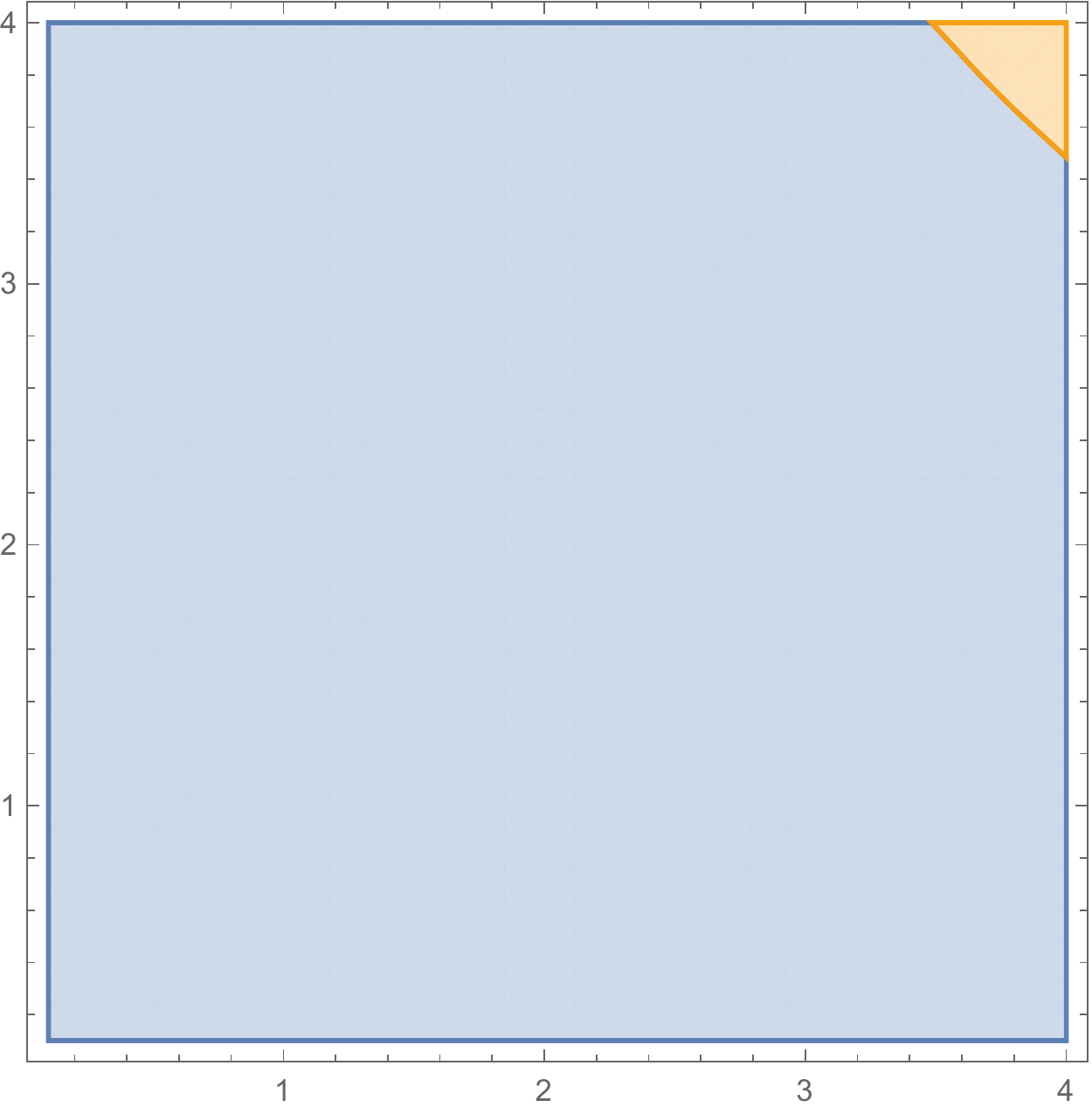} \;\;\;\;
	\includegraphics[scale=0.23]{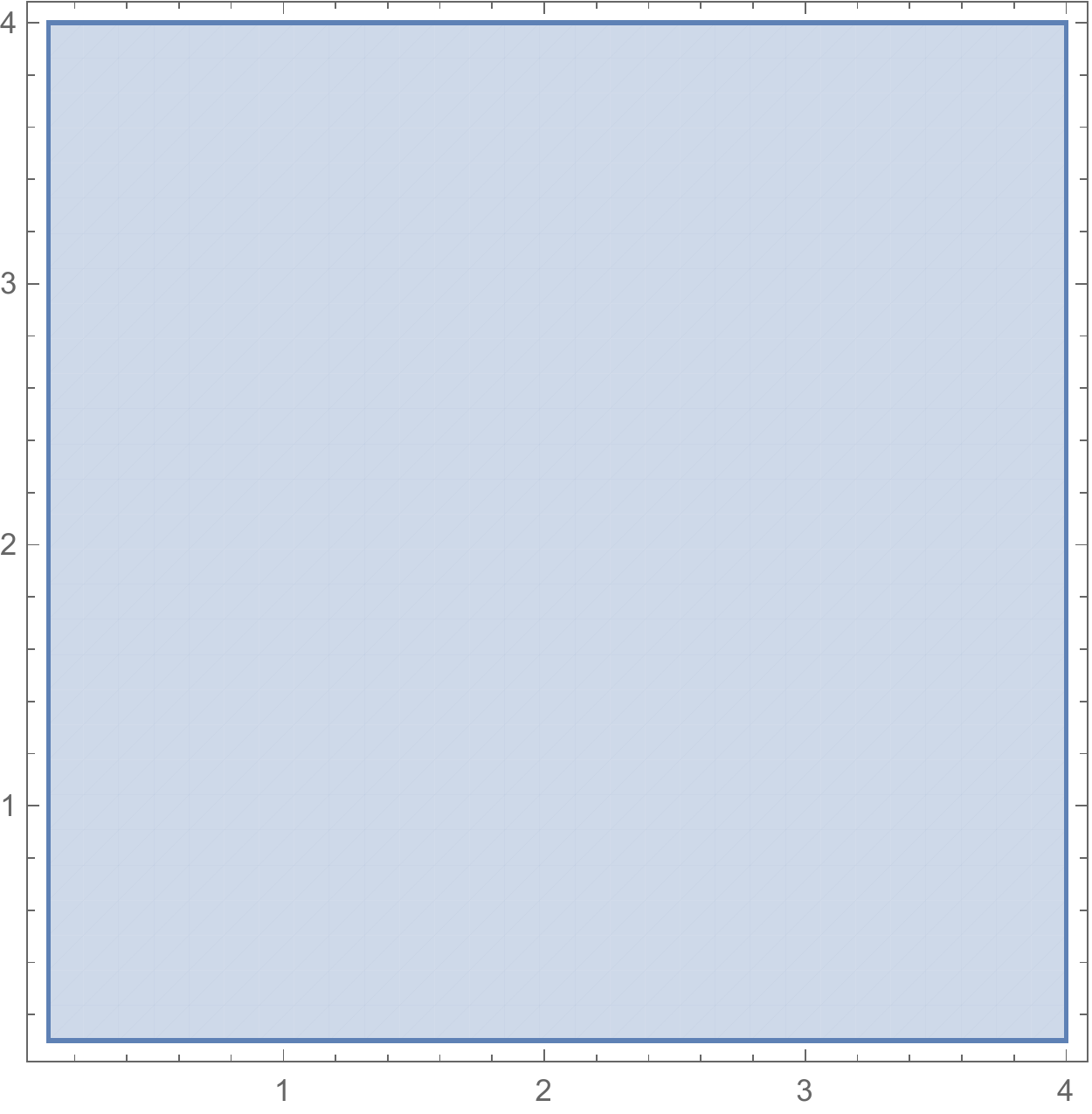}
	\caption{Phase diagrams for HEE in Vaidya-AdS. $z_H=4,~z_h=1,~\ell,m \in(0, 4],~n=3,~x=y=1$, $t=0, 1, 2, 3, 4, 5$. Light purple color corresponds to $(A)\|(B)\|(C)$, light green -- $(A)\|(B,C)$, light blue -- $(A,B)\|(C)$, orange -- $(A,B,C)$.}
	\label{fig:ph-diag-lm}
\end{figure}

It is interesting to note that the same is also true for non-static case.
Phase diagrams for holographic entanglement entropy in the Vaidya--AdS background are presented in  Fig.\ref{fig:ph-diag-xy} and Fig.\ref{fig:ph-diag-lm}.

\newpage
\vspace{10ex} $ $
\subsection{Discontinuity of the holographic entanglement entropy of composite systems}

We study how the holographic entanglement entropy changes if we divide the system into parts. We fix the
total length of the system and divide it into simple or composite parts with distances between them. Holographic entanglement entropy for the disconnected regions was previously studied in \cite{Hubeny:2013hz,Alishahiha:2014jxa,Ben-Ami:2014gsa}. Here, we study specific features of
the case where the total system size is fixed

\subsubsection{2 segments}

\begin{figure}[ht!]
	\centering
	\includegraphics[scale=0.35]{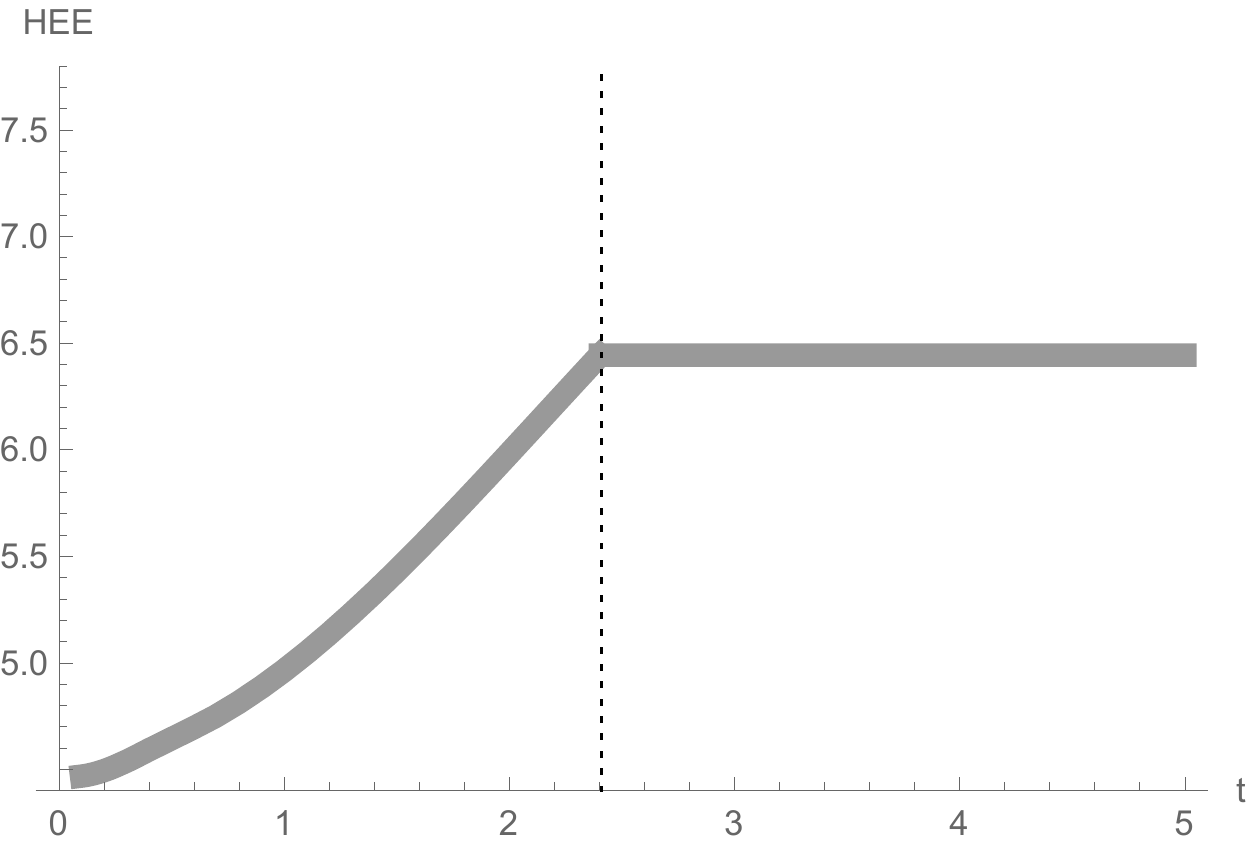}A \;\;\;\;\;\;
	\includegraphics[scale=0.35]{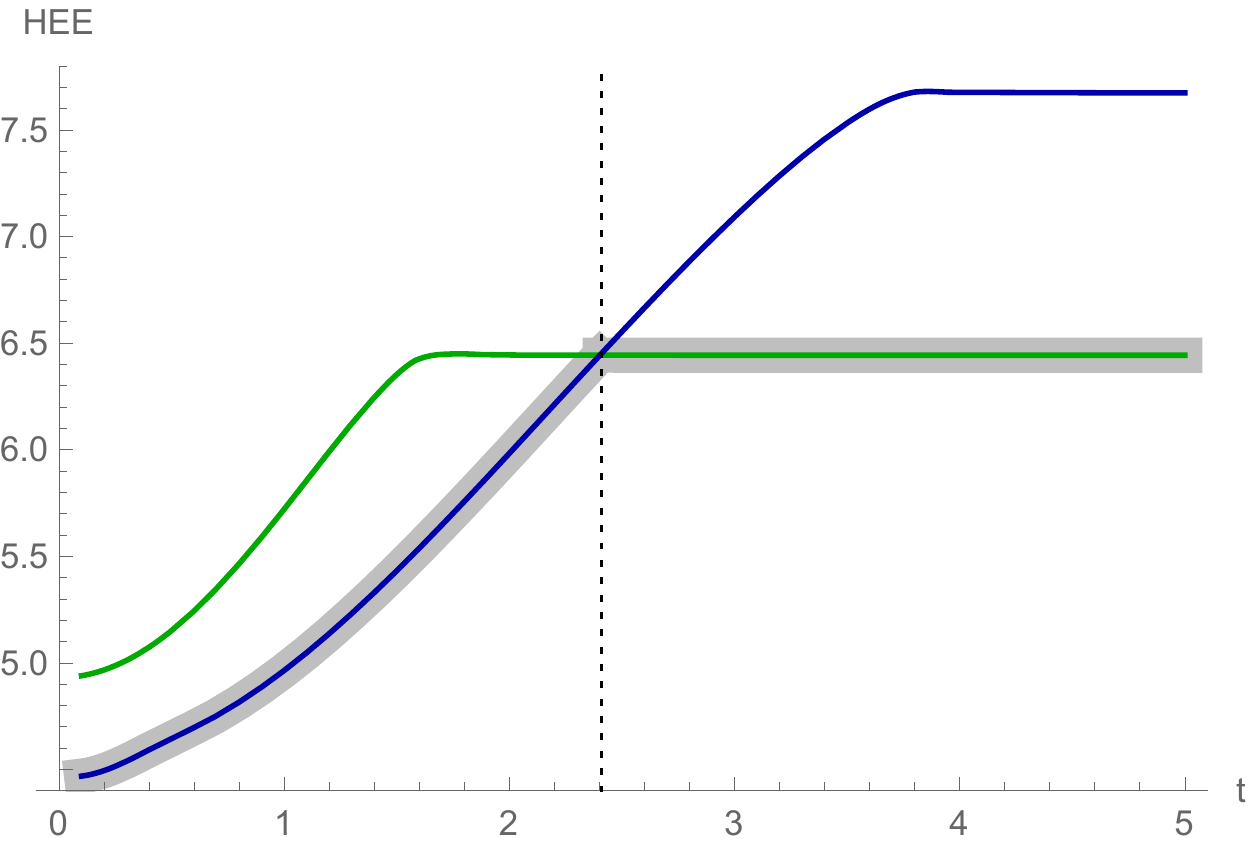}B
	\caption{The holographic entanglement entropy of two segments with one singular point (indicated by dotted vertical line). Blue curve corresponds to connected configuration $(A,B)$, green curve corresponds to disjoint configuration $(A)\|(B)$. In both plots $z_H=3.4$, $z_h=1$, $l=m=3.3$, $x=1$.}
	\label{fig:HEE-2s}
\end{figure}

Let the system consist of two parts each of which is a segment. On Fig.\ref{fig:HEE-2s}.A we plot the holographic entanglement entropy of this system, on Fig.\ref{fig:HEE-2s}.B we also plot connected and disjoint configurations. From these plots one can see that the holographic entanglement entropy has one singular point where configurations change one another. This well known fact has been noted in previous papers \cite{Alishahiha:2014jxa,Ben-Ami:2014gsa}.

\subsubsection{3 segments}
\begin{figure}[ht!]
	\centering
	\includegraphics[scale=0.45]{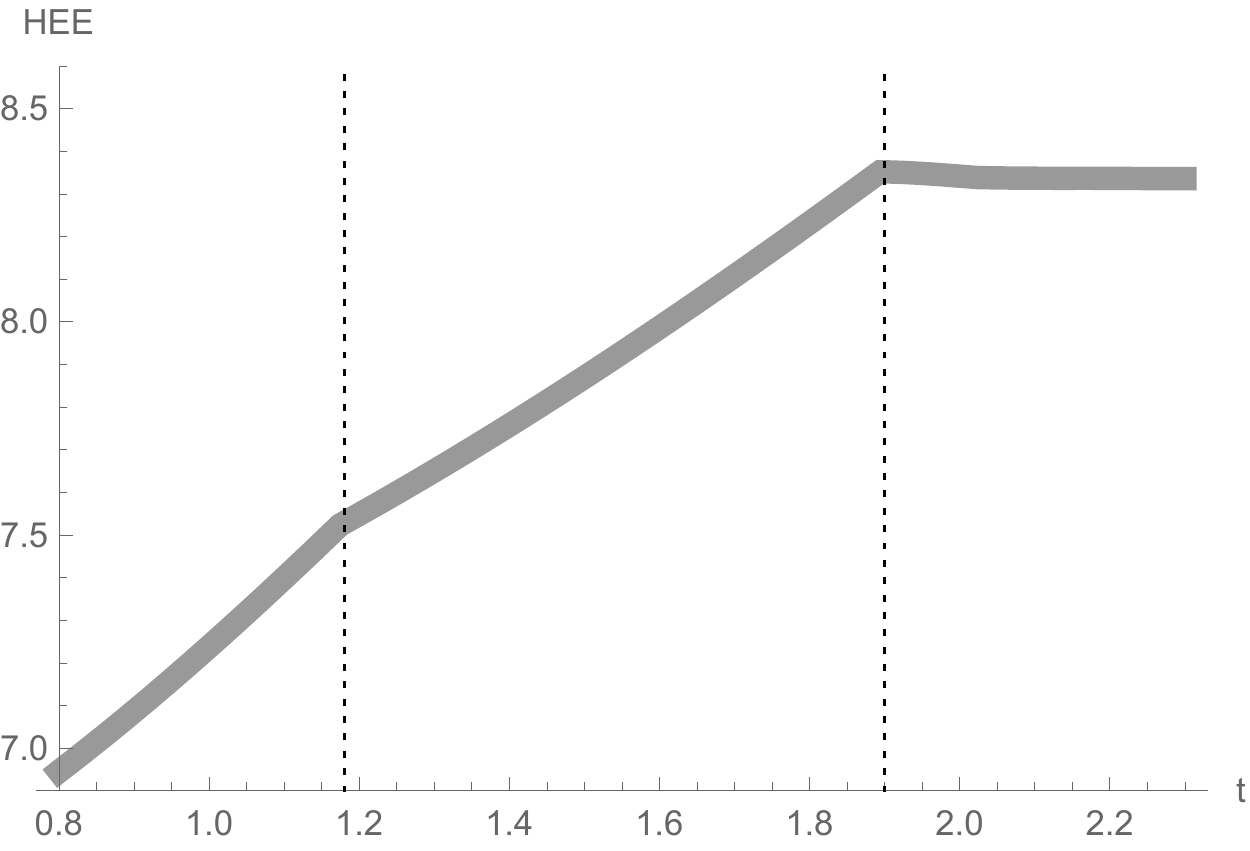}A \;\;\;\;\;\;
	\includegraphics[scale=0.45]{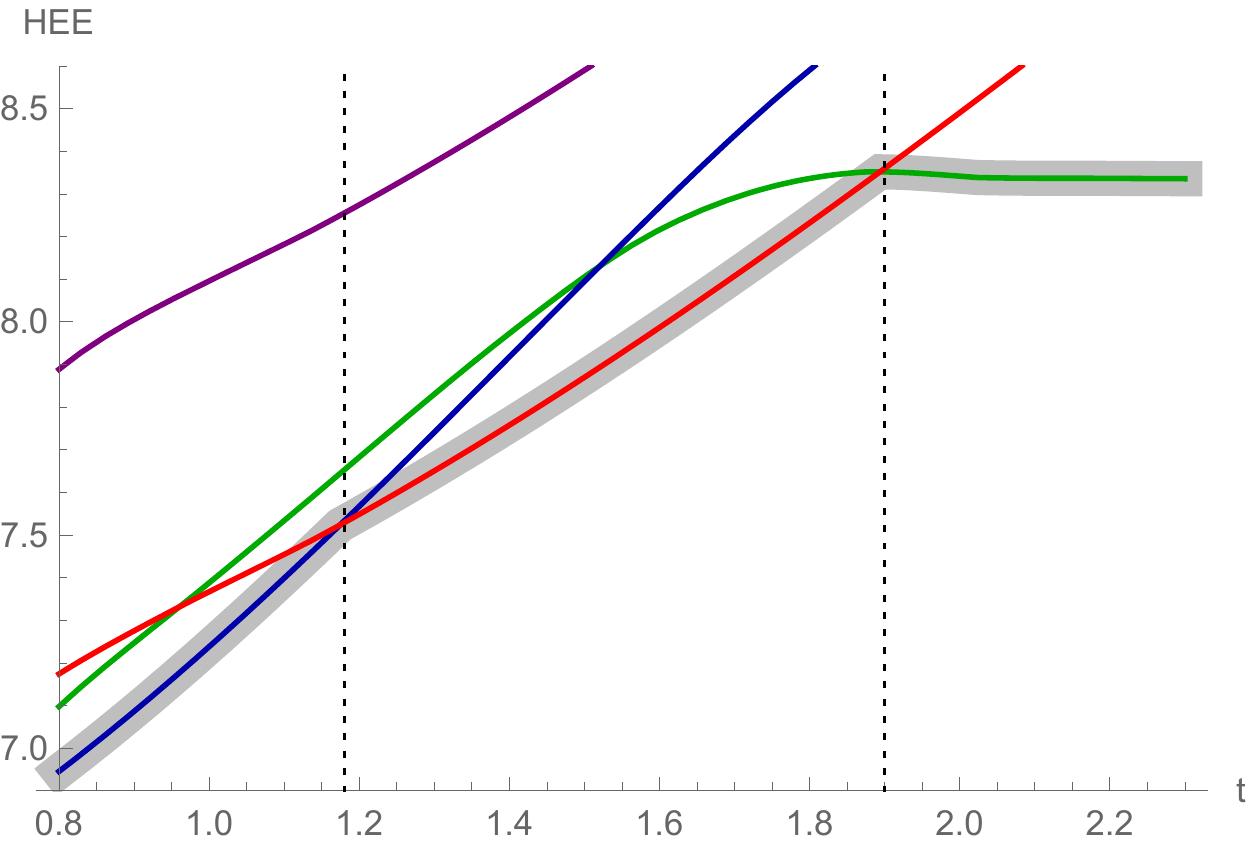}B
	\caption{A. The holographic entanglement entropy with two  singular  points.
		B. The holographic entanglement entropy corresponding  to the first four configurations represented in Fig.\ref{fig:5-config}. Blue curve correspond to $(A,B)\|(C)$, red curve correspond to $(A,B,C)$, green curve correspond to $(A)\|(B)\|(C)$, purple curve correspond to $(A)\|(B,C)$. On both plots $z_H=4.46$, $l=1.77$, $x=0.93$, $m=3.33$, $y=1.87$, $n = 3.76$.}
	\label{fig:HEE-3s-disc-1}
\end{figure}

We now consider a system consisting of three segments
On Fig.\ref{fig:HEE-3s-disc-1}.A we plot the holographic entanglement entropy (thick curve) with two singular points (marked by vertical dashed lines) where configurations change. To see this fact clearly on Fig.\ref{fig:HEE-3s-disc-1}.B we plot the same HEE as on Fig.\ref{fig:HEE-3s-disc-1}.A (thick gray curve) and four curves corresponding to the first four configurations represented on Fig.\ref{fig:5-config}. We can see that in different moments HEE coincides with one of those curves and at the two singular points (marked by vertical dashed lines) these curves change.

\subsubsection{n segments}

It follows from the explicit form of the entanglement entropy evolution that if we divide a segment of length $\ell$ into $n$ parts, calculate the entropy of each part, and multiply it by $n$, then we obtain a value less than the entropy of the whole system: $nS(\ell/n,t)<S(\ell,t)$ for $n>1$. 
Small distances between segments reduce the entanglement entropy, but there is a distance $x_{\mathrm{cr}}$ such that
the entanglement entropy does not change for at least some time $t<t_0$, i.e.,
\begin{equation*}
S(\ell,t)\approx S((\ell-x_{\mathrm{cr}})/2,x_{\mathrm{cr}},(\ell-x_{\mathrm{cr}})/2,t), \;\;\; t<t_0.
\end{equation*}

\vspace{5ex}
\section{Mutual Information of Composite Systems}

\subsection{Appearance of the bell-type mutual information}

For the system consisting of two equal segments with lengths $\ell$ and distance $x$ between them the formula for mutual information reads
\begin{equation*}
I=2\,S(\ell,t)-[S(2\ell+x,t)+S(x,t)].
\end{equation*}
From this formula, we can conclude that if $2\,S(\ell,t)=S(2\ell+x,t)+S(x,t)$ for two different values of $t$, then the plot of the mutual information has a bell shape.

\begin{figure}[!ht]
	\centering
	\includegraphics[scale=0.4]{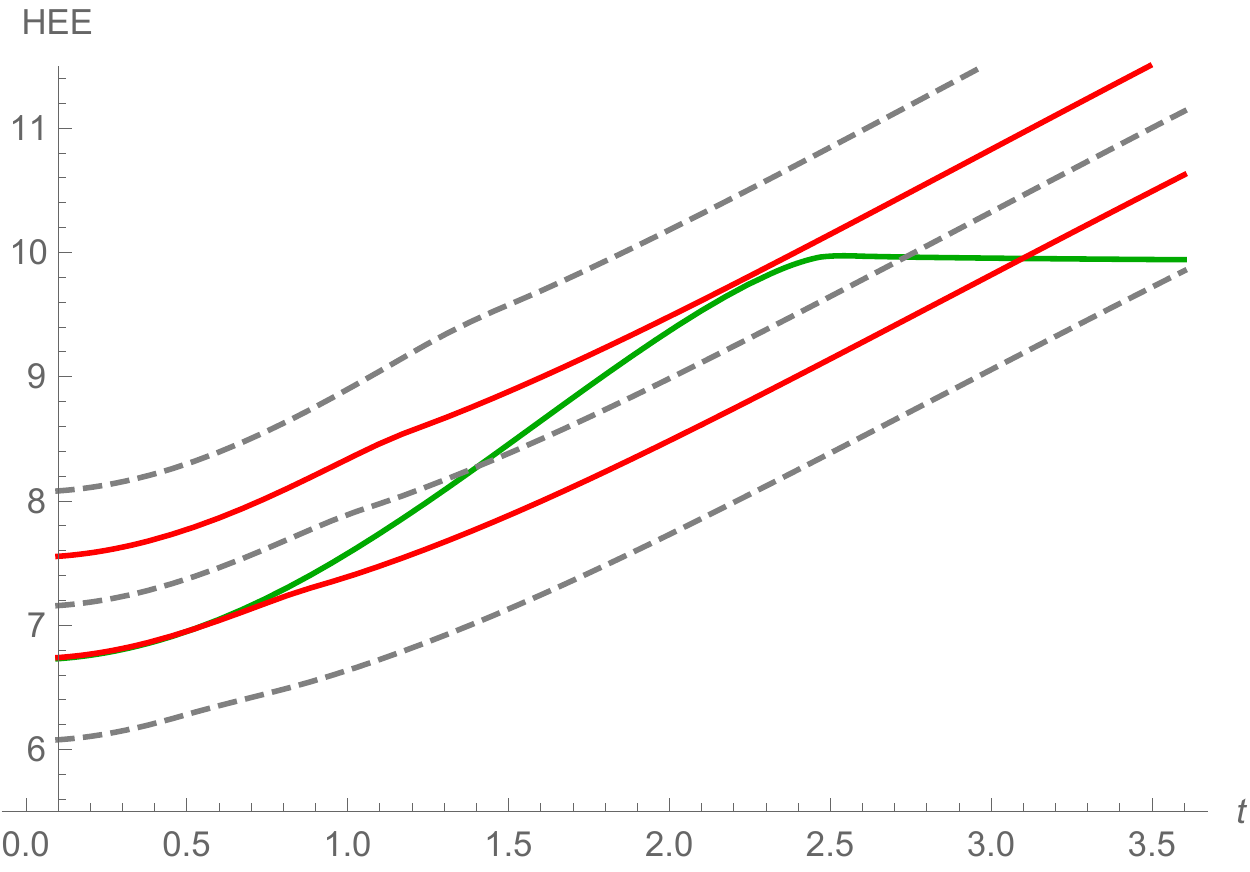}A \;\;\;\;
	\includegraphics[scale=0.53]{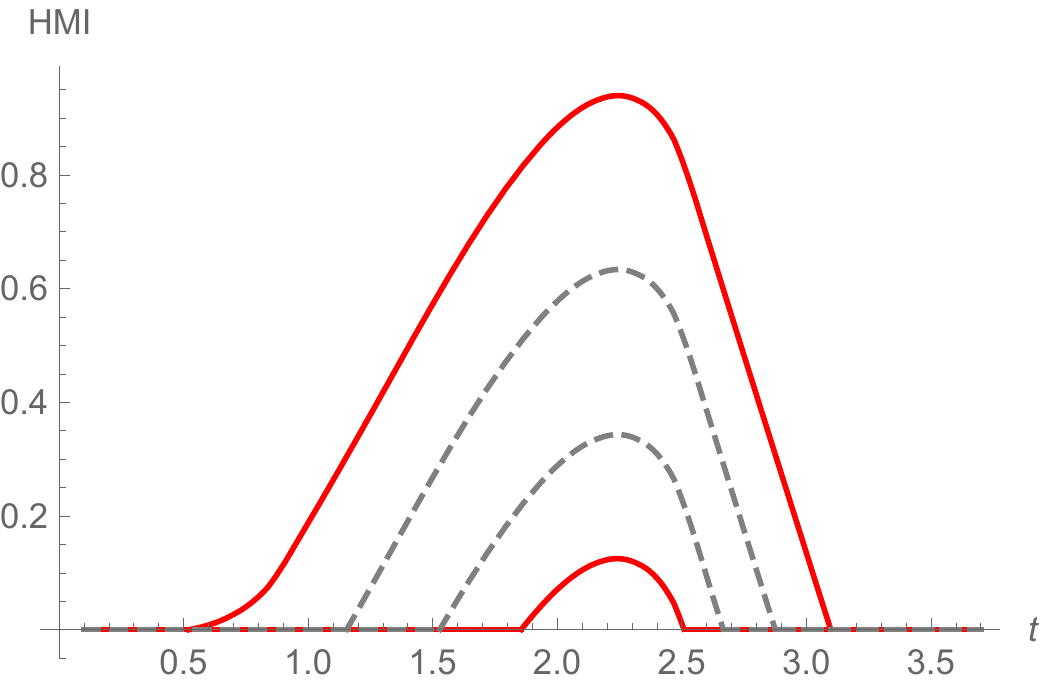}B	
	\caption{The bell-shape mutual information appears between two red curves. On both plots $z_H=3.8,~z_h=1,~\ell_1=\ell_2=5,~x\in [1.7, 2.3]$.}
	\label{fig:2s-bell-origin}
\end{figure}

On Fig.\ref{fig:2s-bell-origin}.A we show the appearance of the bell-type time dependence of the holographic mutual information for two equal segments. 
We can see that the curve representing $2S(\ell,t)$ (green curve) may cross the curves representing $S(2\ell+x,t)+S(x,t)$ (dotted lines) for $x\in [1, 3]$. 
For $x<1.7$ (the lower red curve) there is only one intersection, for $x\in[1.7, 2.3]$ (between two red curves) there are two cross-sections and for $x>2.3$ (the upper red curve) there is no intersection. The double intersection of these two lines corresponds to an appearance of the bell-shaped dependence. 
The lower red curve corresponds to the minimal value of distance $x$ and the upper red curve -- maximal, for which there is the bell-type time dependence of the holographic mutual information. Corresponding mutual information is shown in Fig.\ref{fig:2s-bell-origin}.B.

Note that for these times and length scales $S(2\ell+x,t)$ and $S(x,t)$ have linear time dependence and due to this the difference $2S(\ell)-(S(2\ell+x,t)+S(x,t))$ has a maximum.

\vspace{5ex}
\subsection{Holographic mutual information of 1+2 system}

\subsubsection{Case I}

\begin{figure}[ht!]
	\centering
	\includegraphics[scale=0.45]{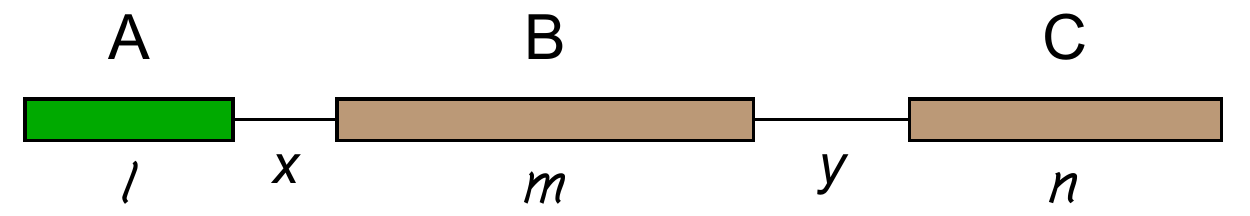}
	\caption{1--2-segment system, case I.}
	\label{fig:Asym}
\end{figure}

In this case, the first (simple) subsystem is a segment $A$ of length $\ell$, the second (composite) subsystem is a union of two segments $B$, $C$ with lengths $m$, $n$ respectively located on one side of segment $A$; in this system $x$, $y$ are distances between segments $A$, $B$ and segments $B$, $C$ respectively (Fig.\ref{fig:Asym}).

\begin{figure}[ht!]
	\centering
	\includegraphics[scale=0.55]{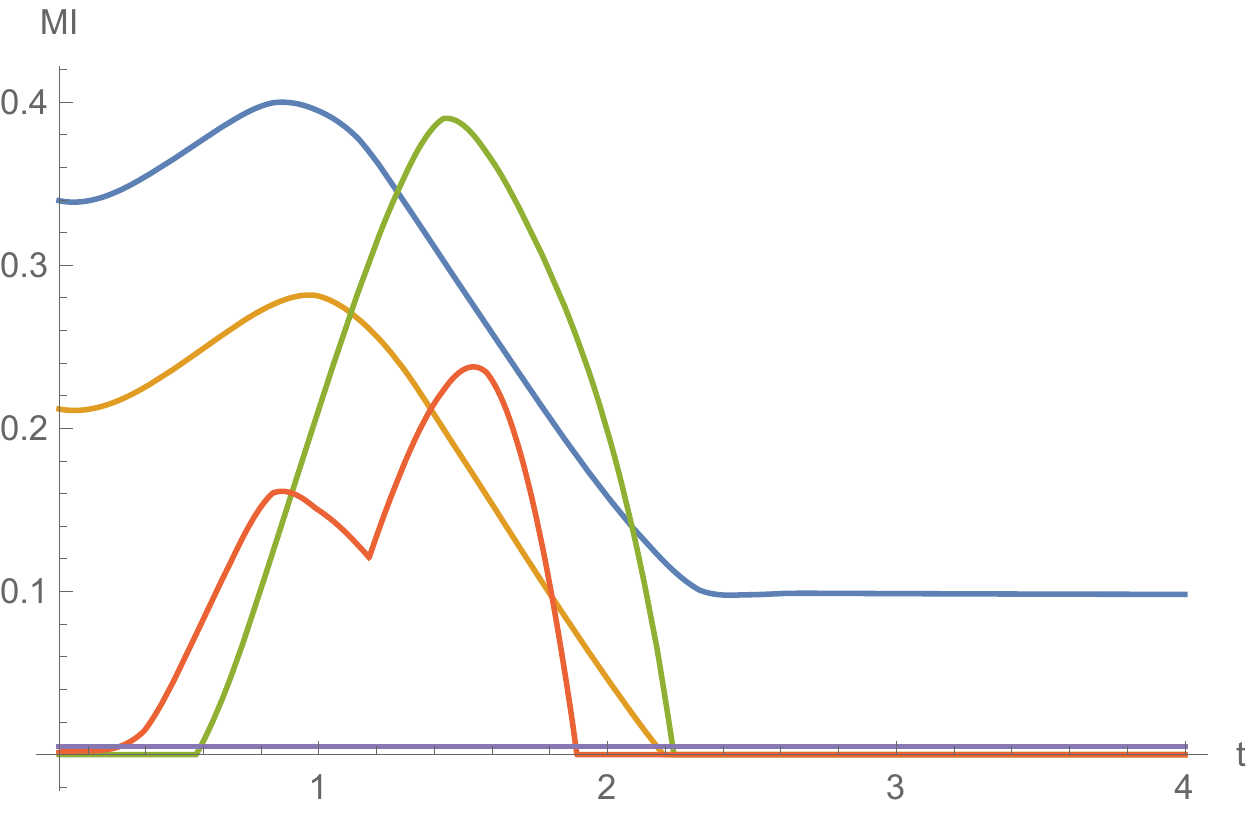}
	\caption{Five types of holographic mutual information behavior for the three-segment system, case I.}
	\label{fig:mi-5forms}
\end{figure}

We have found that the behavior of the holographic mutual information has five types (Fig.\ref{fig:mi-5forms}):

1) Wake-up and scrambling times are absent, and the holographic mutual information is always positive (blue curve);

2) Wake-up time is absent, but scrambling time is present (orange curve);

3) Wake-up and scrambling times are present, and a plot of the holographic mutual information has a
bell shape (green curve);

4) Wake-up and scrambling times are present, and a plot of the holographic mutual information has a
two-hump shape (red curve);

5) The holographic mutual information is identically equal to zero (purple curve).

\begin{figure}
	\centering
	\includegraphics[scale=0.45]{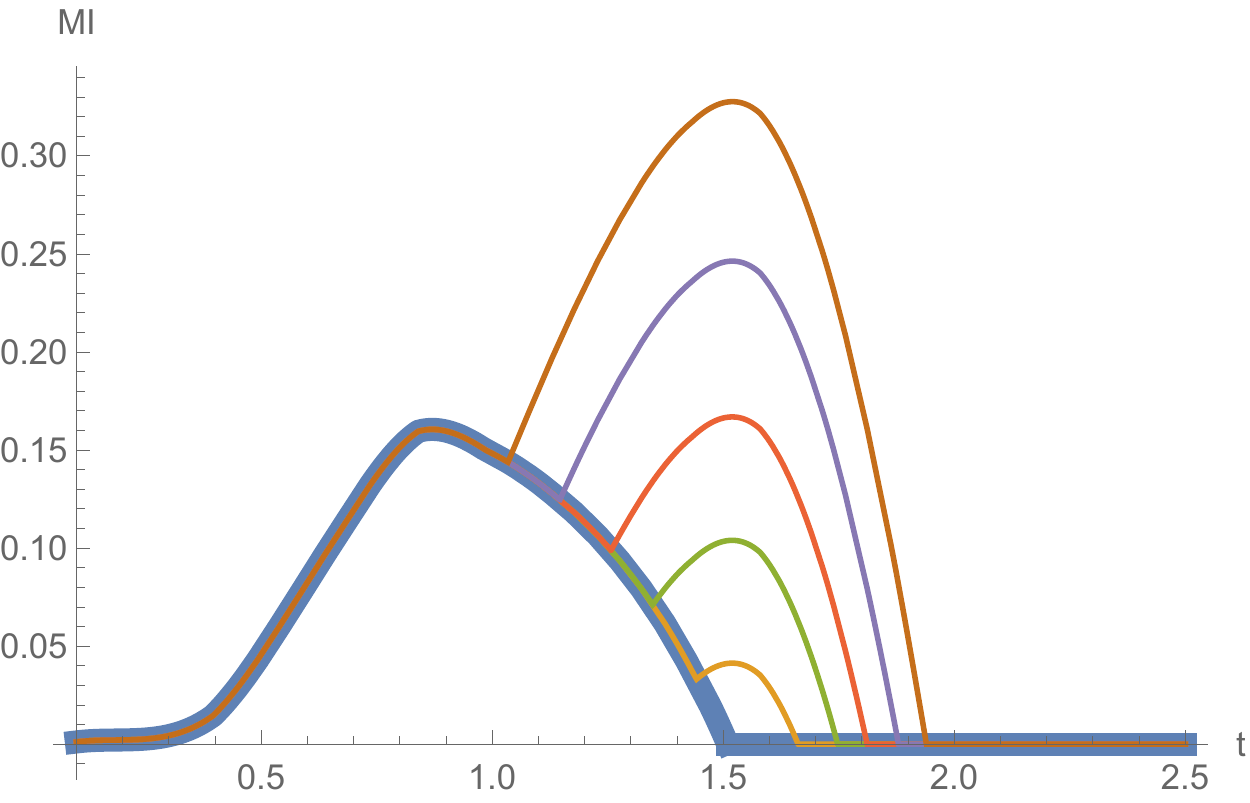}A \;\;\;\;
	\includegraphics[scale=0.45]{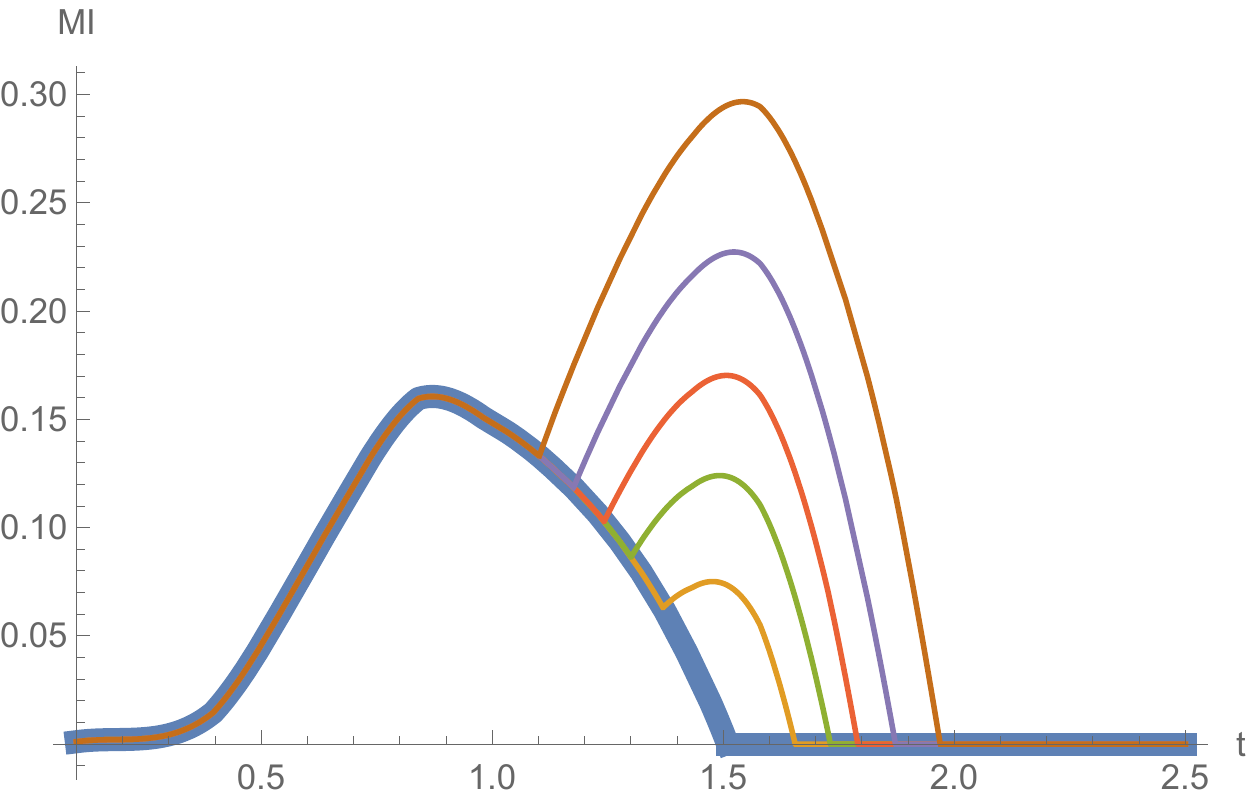}B
	\caption{Appearance of a two-humped form of the HMI. \textbf{A.} $z_H=4.5$, $z_h=1$, $\ell=1.77$, $m=3.3$, $n=3.7$, $x=0.93$, $y=1.98, 1.94, 1.90, 1.85, 1.80$. \textbf{B.} $z_H=4.5$, $z_h=1$, $\ell=1.77$, $m=3.3$, $n=3.4, 3.5, 3.6, 3.7, 3.9$, $x=0.93$, $y=1.87$.}
	\label{fig:HMI-3s-Asymm}
\end{figure}

Let $A$ and $B$ be two segments such that the corresponding holographic mutual information has a bell shape. Fig.\ref{fig:HMI-3s-Asymm}.A shows an appearance of the second hump as segment $C$ approaches segments $A$ and $B$ from the right (as $y$ decreases). Fig.\ref{fig:HMI-3s-Asymm}.B shows an appearance of the second hump as the length of segment $C$ increases with the distances fixed (as $n$ increases). 

\begin{figure}[ht!]
	\centering
	\includegraphics[scale=0.29]{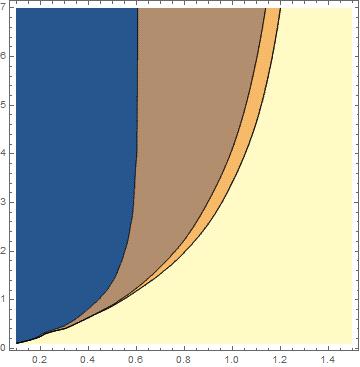}A \;\;\;\;\;\;\;\;
	\includegraphics[scale=0.30]{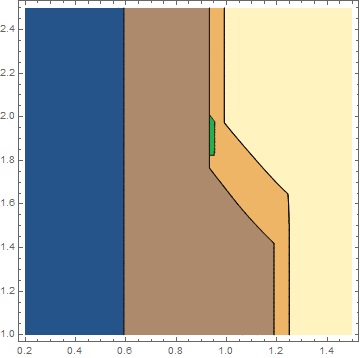}B	
	\caption{Zones of different behavior of the HMI. Blue color corresponds to $HMI>0$, brown -- presence of scrambling time, orange -- bell shape, green -- two-hump shape, yellow -- $HMI\equiv0$. \textbf{A.} The system consists of two segments, horizontal axis corresponds to $x$, vertical -- to $m$, $z_H=4.46$, $z_h=1$, $\ell=1.77$ and $x, m$ are varied. \textbf{B.} The system consists of three segments, horizontal axis corresponds to $x$, vertical -- to $y$, total length of the system is 12, $z_H=4.46$, $z_h=1$, $\ell=1.77$, $m=3.33$, $n=12-1.77-3.33-x-y$ and $x, y$ are varied.}
	\label{fig:2slmn-zones}
\end{figure}

In Fig.\ref{fig:2slmn-zones}.B, we show the zones of different behavior (see the five types above) of the holographic mutual information for the three-segment system depending on x and y. It can be seen from this picture that the two-humped zone corresponds to rather narrow intervals of parameters.

\newpage
\subsubsection{Case II}

\begin{figure}[ht!]
	\centering
	\includegraphics[scale=0.45]{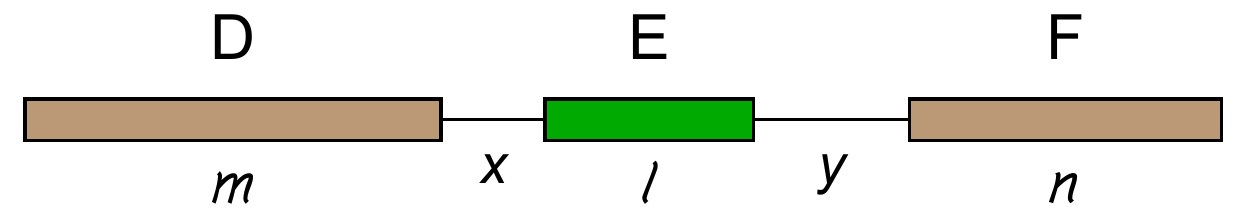}
	\caption{1--2-segment system, case II.}
	\label{fig:Symm}
\end{figure}

In this case, the first (simple) subsystem is a segment $A$ of length $\ell$, the second (composite) subsystem is a union of two segments $B$, $C$ with lengths $m$, $n$ respectively, located on opposite sides of segment $A$; in this system, $x$, $y$ are distances between segments $A$, $B$ and segments $A$, $C$ respectively (Fig.\ref{fig:Symm}).

\begin{figure}[ht!]
	\centering
	\includegraphics[scale=0.55]{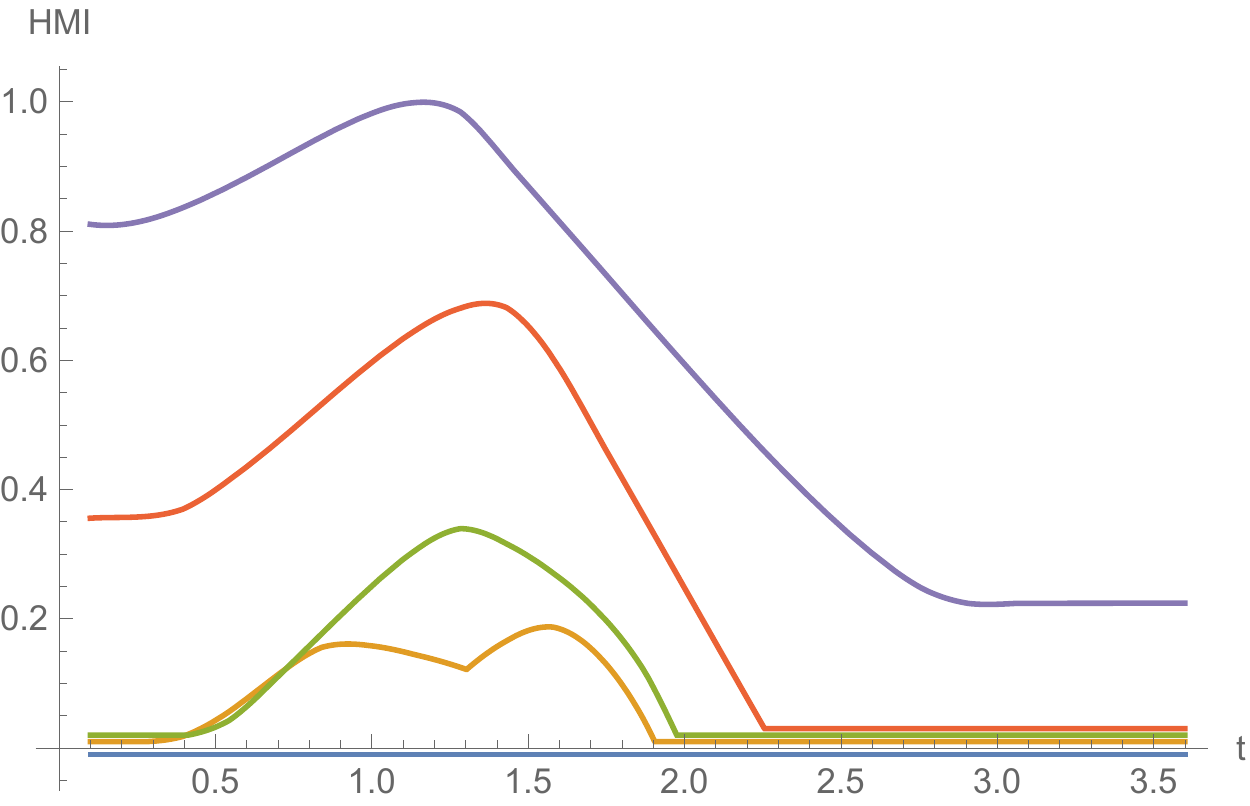}
	\caption{Five types of holographic mutual information behavior for composite system of three segments, case II.}
	\label{fig:hmi-case2-5forms}
\end{figure}

In this case, we have found generally the same typical holographic mutual information behavior as in the case I (Fig.\ref{fig:hmi-case2-5forms}):

1) Wake-up and scrambling times are absent, and the holographic mutual information is always positive (purple curve);

2) Wake-up time is absent, but scrambling time is present (red curve);

3) Wake-up and scrambling times are present, and a plot of the holographic mutual information has a
bell shape (green curve);

4) Wake-up and scrambling times are present, and a plot of the holographic mutual information has a
two-hump shape (orange curve);

5) The holographic mutual information is identically equal to zero (blue curve).

\begin{figure}
	\centering
	\includegraphics[scale=0.45]{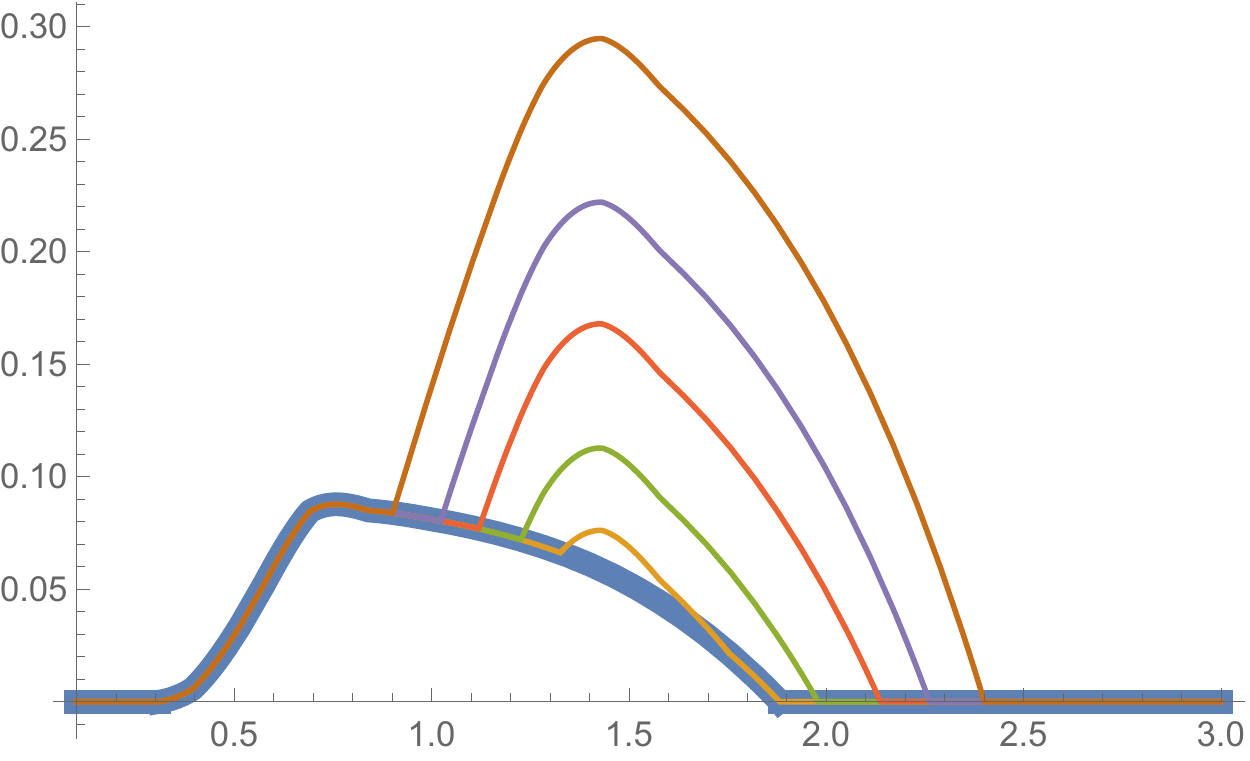}A \;\;\;\;
	\includegraphics[scale=0.45]{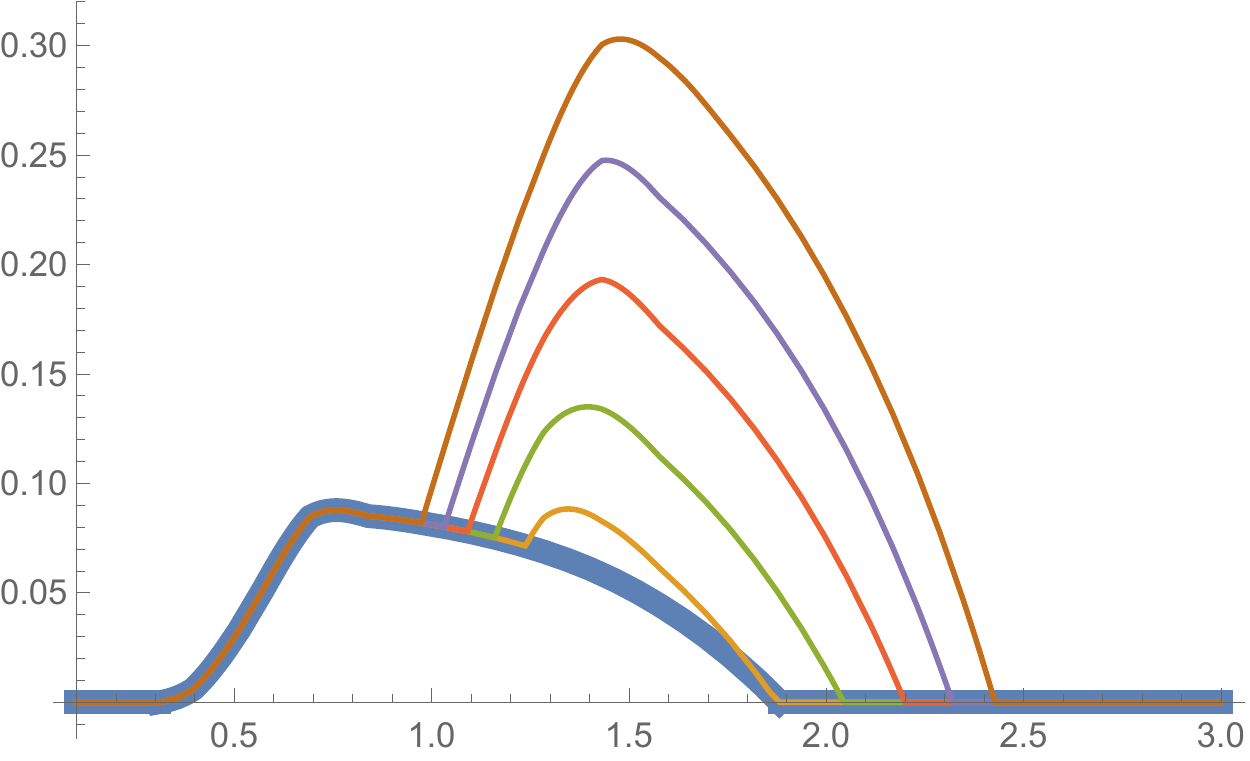}B
	\caption{Appearance of a two-humped form of the HMI. \textbf{A.} $z_H=3.5$, $z_h=1$, $m=2.82$, $\ell=5.45$, $n=1.46$,  $x=1.65, 1.63, 1.60, 1.57, 1.53$, $y=0.924$. \textbf{B.} $z_H=3.5$, $z_h=1$, $m=2.72, 2.78, 2.85, 2.92, 3.00$, $\ell=5.45$, $n=1.46$,  $x=1.6$, $y=0.924$.}
	\label{fig:HMI-3s-Symm}
\end{figure}

Let $E$ and $F$ be two segments such that the corresponding holographic mutual information has a bell shape. Fig.\ref{fig:HMI-3s-Symm}.A shows an appearance of the second hump as segment $D$ approaches segments $E$ and $F$ from the left (as $x$ decreases). Fig.\ref{fig:HMI-3s-Symm}.B shows an appearance of the second hump as the length of segment $D$ increases with the distances fixed (as $m$ increases).

\begin{figure}[ht!]
	\centering
	\includegraphics[scale=0.32]{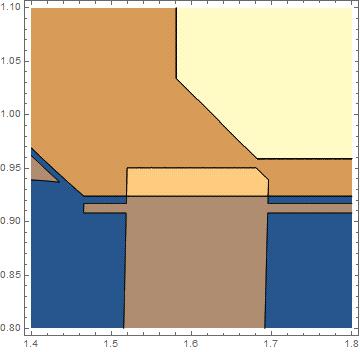}
	\caption{Zones of different behavior of HMI. The system consists of three segments. Horizontal axis corresponds to $x$, vertical -- to $y$. $z_H=3.5$, $z_h=1$, $\ell=5.48$, $m=2.9$, $n=1.46$ and $x, y$ are varied. Yellow zone corresponds to $HMI\equiv0$, orange zone -- bell-type form, darker yellow -- two-humped form, brown zone corresponds to the presence of scrambling time and local minimum, blue zone corresponds to the presence of scrambling time.}
	\label{fig:3smln-2hump-zones}
\end{figure}

In Fig.\ref{fig:3smln-2hump-zones}, we show the zones of different behavior (see the five types above) of the holographic mutual information for the three-segment system depending on x and y. It can be seen from this picture that the two-humped zone corresponds to rather narrow intervals of parameters.

\subsubsection{Comparison of Case I and Case II}

It is interesting to compare the holographic mutual information for the two cases. We can do it in two ways.

\begin{figure}[ht!]
	\centering
	\includegraphics[scale=0.4]{Asymm} \;\;\;\;\;\;\;
	\includegraphics[scale=0.4]{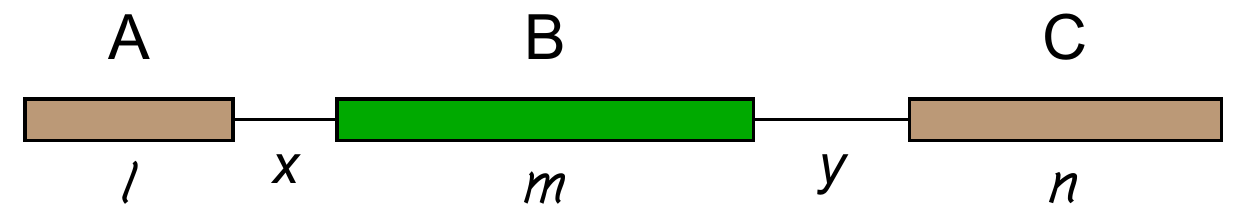}
	\caption{Left plot: $I(A;B \cup C)=S(A)+S(B \cup C)-S(A \cup B \cup C)$, right plot: $I(B;A \cup C)=S(B)+S(A \cup C)-S(A \cup B \cup C)$.}
	\label{fig:asymm}
\end{figure}

\textbf{First Way.} The geometry of the system does not change, but the partition of the system into two subsystems changes (Fig. \ref{fig:asymm})

\begin{figure}[ht!]
	\centering
	\includegraphics[scale=0.32]{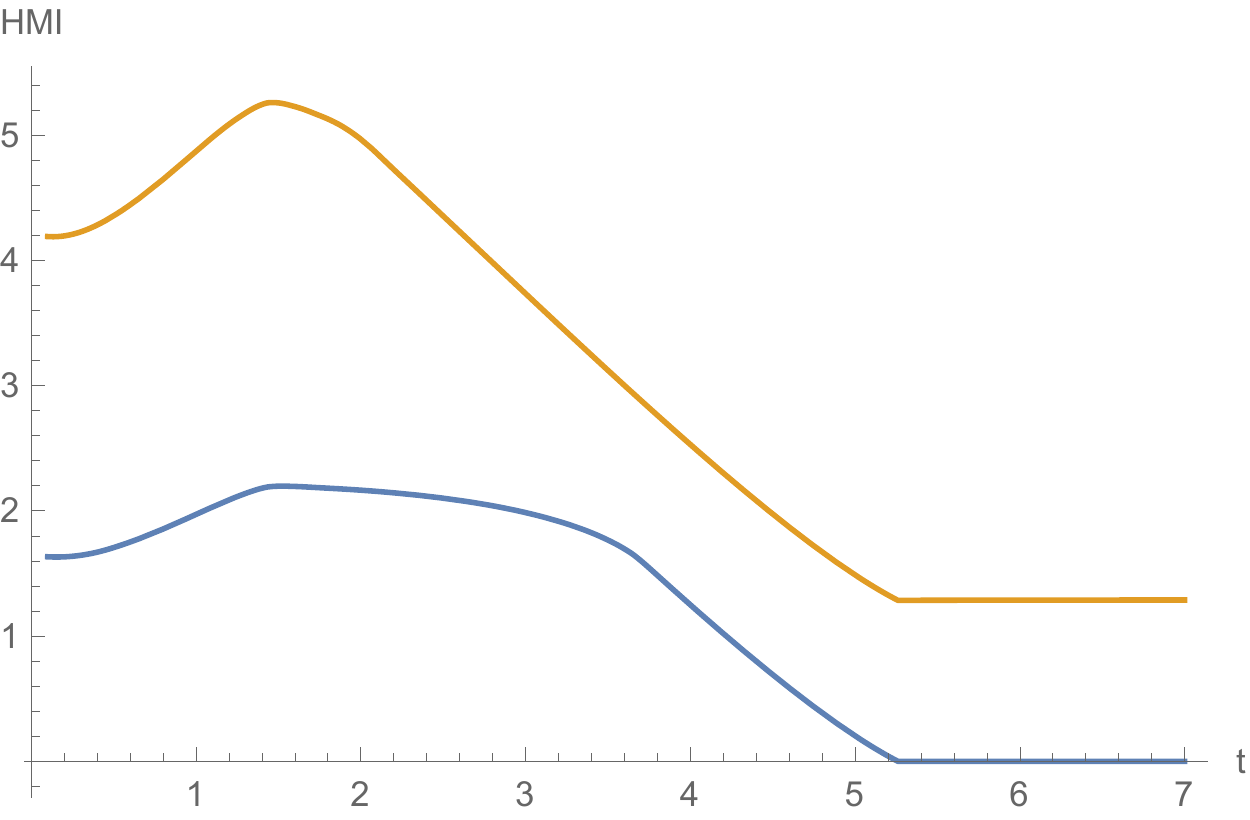}A \;\;\;
	\includegraphics[scale=0.32]{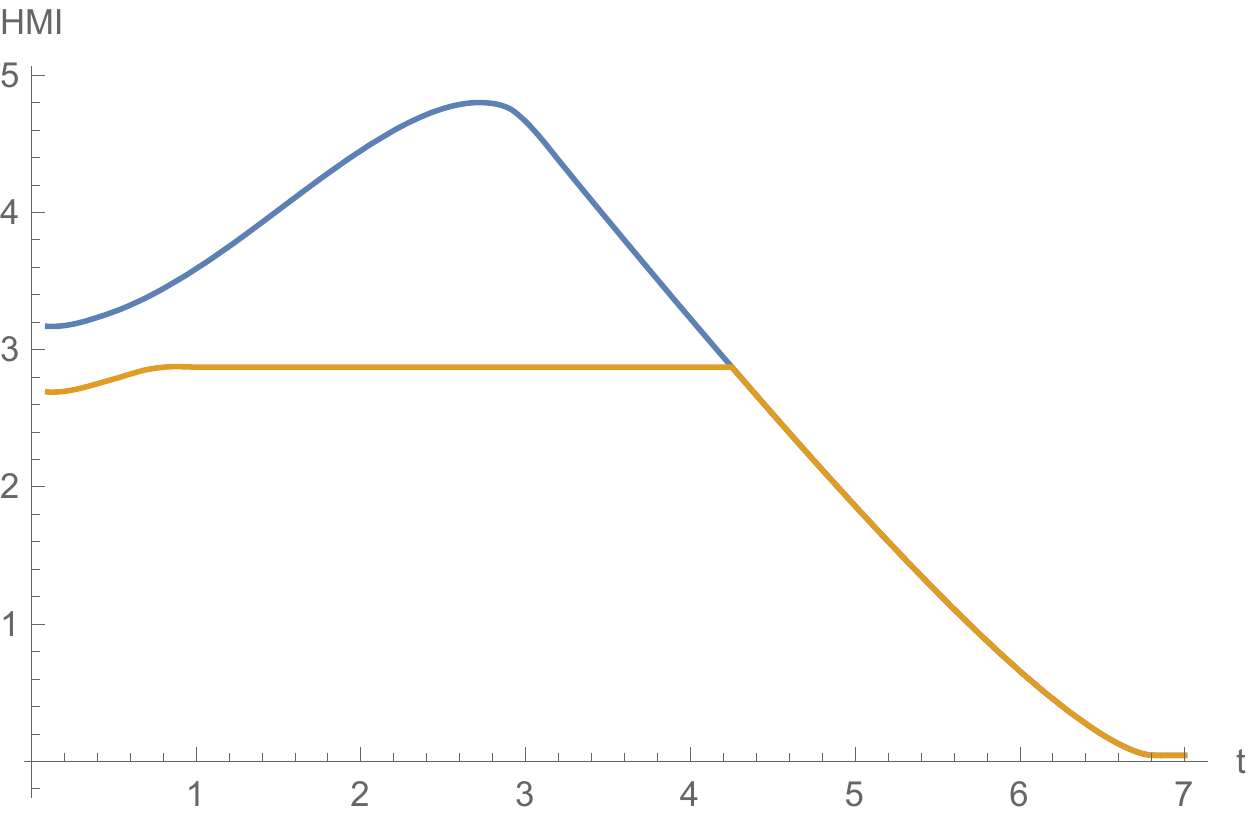}B \;\;\;
	\includegraphics[scale=0.32]{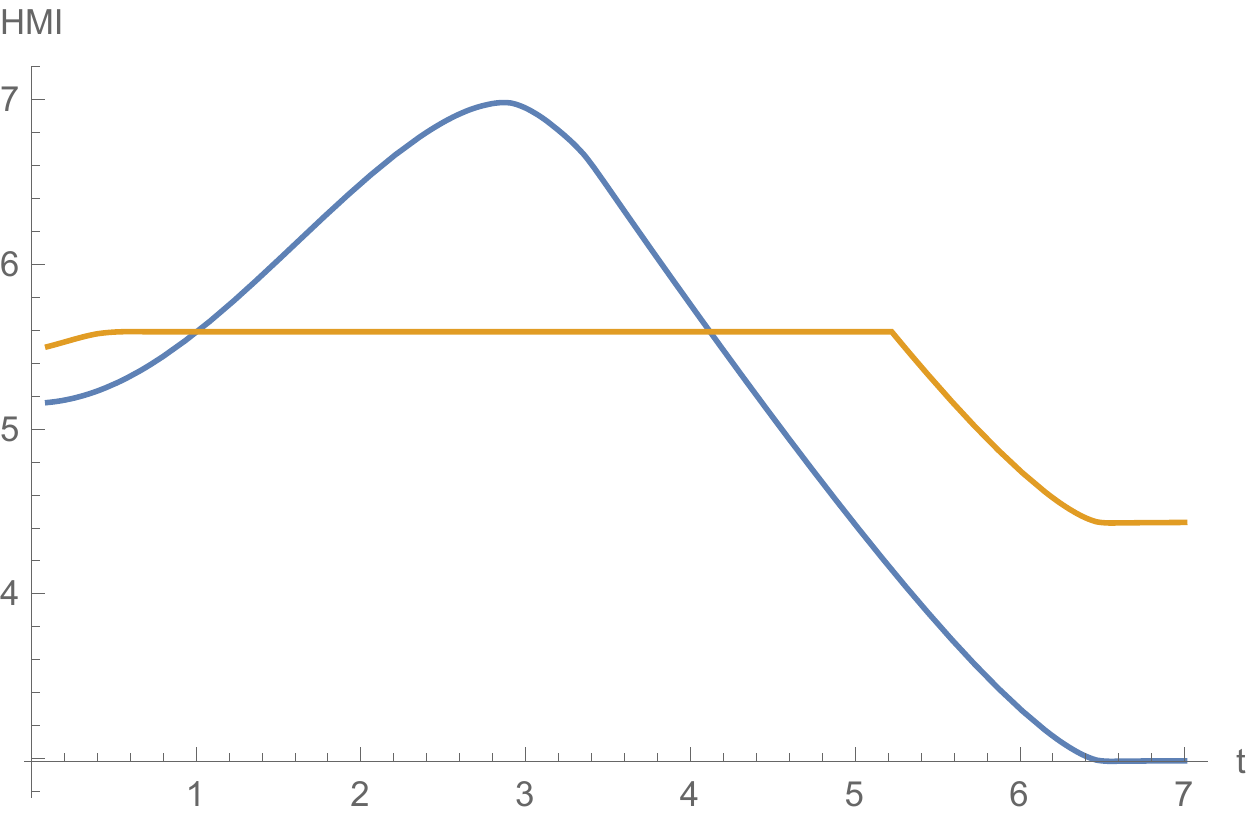}C
	\caption{Blue curve corresponds to $I(A;B\cup C)$, orange curve corresponds to $I(B;A\cup C)$. \textbf{A.} $z_H=3,~ z_h=1,~ l=3,~ x=0.7,~ m=4,~ y=0.4,~ n=3$. \textbf{B.} $z_H=4,~ z_h=1,~ l=6,~ x=0.5,~ m=0.6,~ y=0.5,~ n=6$. \textbf{C.} $z_H=4,~ z_h=1,~ l=6,~ x=0.2,~ m=0.6,~ y=0.2,~ n=6$.}
	\label{fig:ABC-BAC}
\end{figure}

In this case, all three situations are possible:
\begin{equation}
I(A;B\cup C) < I(B;A\cup C),~~ I(A;B\cup C) > I(B;A\cup C),~~ I(A;B\cup C) = I(B;A\cup C)
\end{equation}
for some $A, B, C$ and at different instants $t$ (Fig.\ref{fig:ABC-BAC}).

\begin{figure}[ht!]
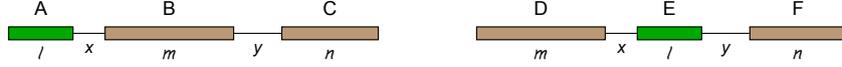

	\centering
	\includegraphics[scale=0.4]{Asymm} \;\;\;\;\;\;\;
	\includegraphics[scale=0.4]{Symm2}
	\caption{Left plot: $I(A;B \cup C)=S(A)+S(B \cup C)-S(A \cup B \cup C)$, right plot: $I(E;D \cup F)=S(E)+S(D \cup F)-S(D \cup E \cup F)$.}
	\label{fig:symm}
\end{figure}

\textbf{Second Way.} The geometry of the system changes, but the partition of the system into two subsystems
does not change. (Fig. \ref{fig:symm})

\begin{figure}[ht!]
	\centering
	\includegraphics[scale=0.32]{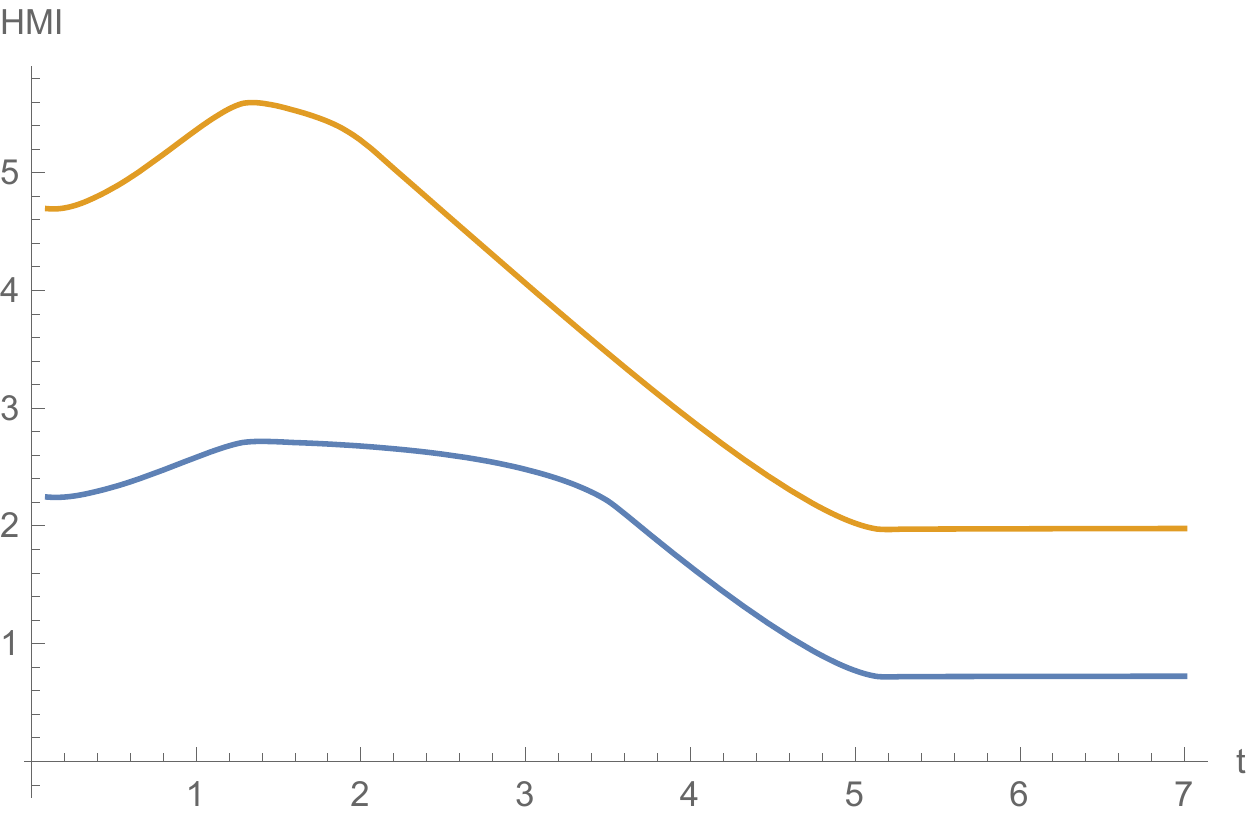}A \;\;\;
	\includegraphics[scale=0.32]{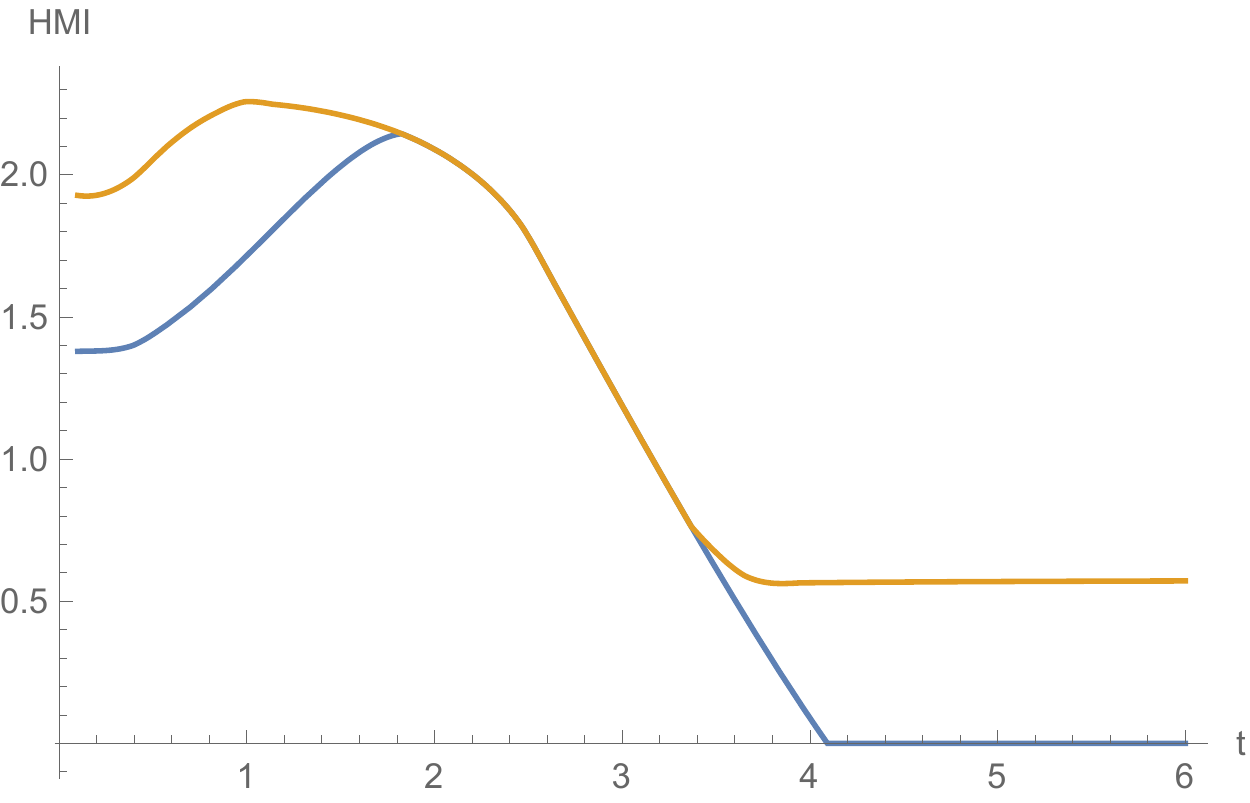}B \;\;\;
	\includegraphics[scale=0.32]{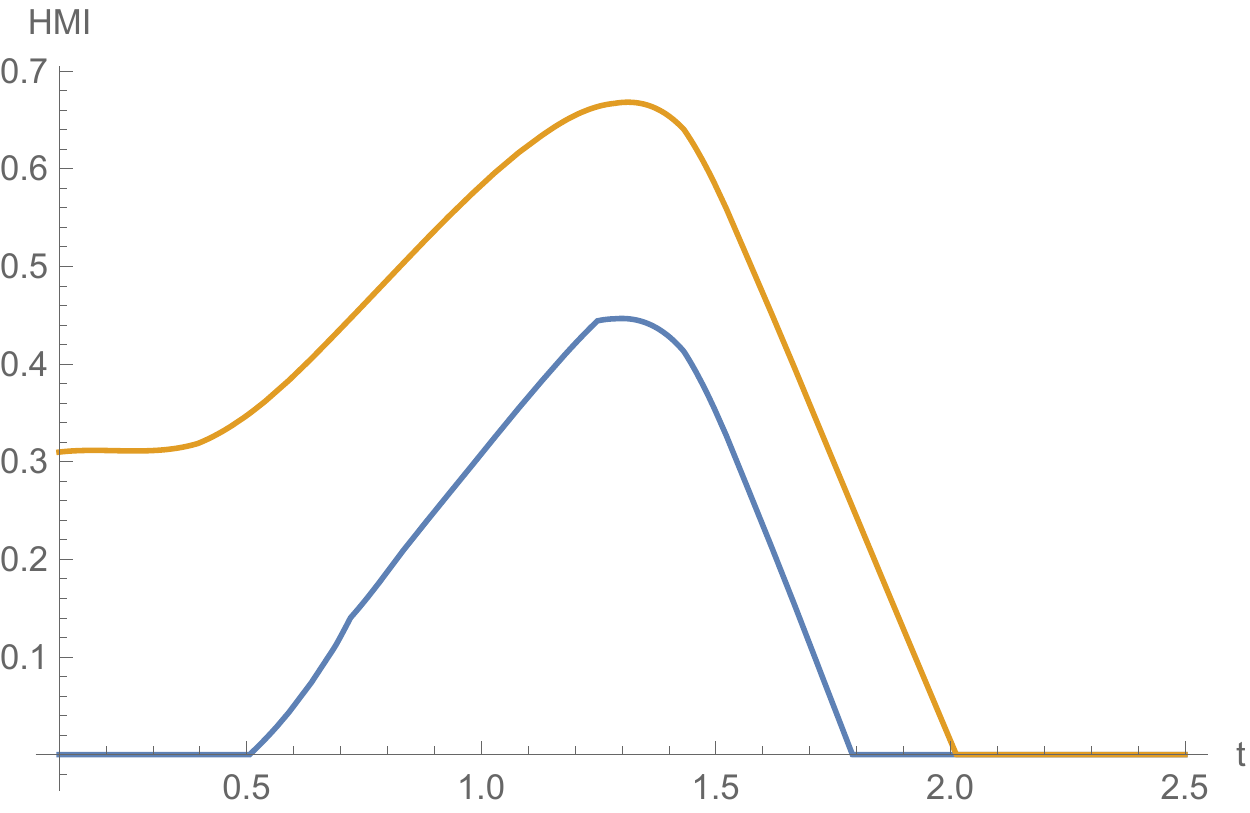}C
	\caption{Blue curve corresponds to $I(A;B\cup C)$, orange curve corresponds to $I(E;D\cup F)$. \textbf{A.} $z_H=3,~ z_h=1,~ l=2.7,~ x=0.5,~ m=2.7,~ y=0.4,~ n=4$. \textbf{B.} $z_H=3,~ z_h=1,~ l=5,~ x=0.8,~ m=1.3,~ y=0.5,~ n=2$. \textbf{C.} $z_H=4,~ z_h=1,~ l=2.9,~ x=1.4,~ m=2,~ y=1,~ n=3$.}
	\label{fig:ABC-DEF}
\end{figure}

In this case, we numerically find that the following inequality is satisfied
\begin{equation}
I(A;B \cup C) \leqslant I(E;D \cup F)
\end{equation}
for any $A, B, C$ and for any time $t$ (Fig.\ref{fig:ABC-DEF}).

\vspace{3ex}
\section{Conclusion}

Using the holographic approach, we studied the time evolution of the mutual information of a composite system during the heating process and showed how the mutual information between two sub-systems can be controlled by changing the geometric system parameters. We have presented a more detailed analysis for a three-segment system consisting of two subsystems, studying an asymmetric type of organization (case I) and a symmetric type of organization (case II). In both cases, we have found 5 types of behavior of the holographic mutual information for a system containing one composite part (two segments) and one simple part (one segment):

1) Wake-up and scrambling times are absent, and the holographic mutual information is always positive;

2) Wake-up time is absent, but scrambling time is present;

3) Wake-up and scrambling times are present, and a plot of the holographic mutual information has a bell shape;

4) Wake-up and scrambling times are present, and a plot of the holographic mutual information has a two-hump shape;

5) The holographic mutual information is identically equal to zero.

We have shown that the holographic mutual information in the symmetric case II is greater or equal to the holographic mutual information in the asymmetric case I (using the second way of comparison). We paid particular attention to finding a special shape of the plot of the holographic mutual information: a double-hump. We have found that this double-hump shape is realized only for quite small ranges of system parameters.

\vspace{3ex}
\section{Acknowledgments}

This work was supported by the Russian Science Foundation (project №17-71-20154).


\vspace{3ex}
\begin{thebibliography}{99}
	\bibitem{OhyaVol} Masanori Ohya, Igor Volovich, \textit{Mathematical Foundations of Quantum Information and Computation and Its Applications to Nano- and Bio-systems}, Springer, 2011.
	
	\bibitem{AreVol-Photo}
	I. Aref'eva and I. Volovich, "Holographic Photosynthesis"; arXiv:1603.09107 [hep-th].
	
	\bibitem{QEB} M. Mohseni, Y. Omar, G. S. Engel, and M. B. Plenio, eds., \textit{Quantum Effects in Biology}, Cambridge Univ. Press, Cambridge (2014).
		
	\bibitem{Aharony:1999ti}
	O.~Aharony, S.~S.~Gubser, J.~M.~Maldacena, H.~Ooguri and Y.~Oz,
	``Large N field theories, string theory and gravity,''
	Phys.\ Rept.\  {\bf 323}, 183 (2000)
	[hep-th/9905111].
	
	\bibitem{IA}   I.~Ya.~Aref'eva,
	``Holographic approach to quark-gluon plasma in heavy ion collisions,''
	Phys.\ Usp.\  {\bf 57}, 527 (2014).
	
	\bibitem{IA15} I. Y. Aref’eva, "QGP time formmation in holographic shock waves model of heavy ion collisions," Theor. Math. Phys., 184, 1239–1256 (2015); arXiv:1503.02185v1 [hep-th] (2015).
		
	\bibitem{DAIA1} D. S. Ageev and I. Ya. Aref’eva, “Waking and scrambling in holographic heating up,” Theor. Math. Phys., 193, 1534-1546; arXiv:1701.07280[hep/th].
	
	
	\bibitem{DAIA2}
	D.~S.~Ageev and I.~Y.~Aref'eva,
	"Memory Loss in Holographic Non-equilibrium Heating,"
	arXiv:1704.07747 [hep-th].
	
	\bibitem{KV} I. V. Volovich, S. V. Kozyrev, "Manipulation of states of a degenerate quantum system", Proc. Steklov Inst. Math., 294 (2016), 241–251.
	
	\bibitem{AKh} I. Ya. Aref'eva, M. A. Khramtsov, "AdS/CFT prescription for angle-deficit space and winding geodesics", JHEP, 2016, no. 4, 121 , 21 pp.
	
	\bibitem{TV}
	A. S. Trushechkin, I. V. Volovich, "Perturbative treatment of inter-site couplings in the local description of open quantum networks", EPL, 113:3 (2016), 30005 , 6 pp.
	
	\bibitem{Vol}	Igor V. Volovich, "Cauchy--Schwarz inequality-based criteria for the non-classicality of sub-Poisson and antibunched light", Phys. Lett. A, 380:1 (2016), 56–58.
	
	\bibitem{AHT} I. Y. Aref’eva, M. A. Khramtsov, and M. D. Tikhanovskaya, “Thermalization after holographic bilocal quench,” JHEP, 2017, 115 (1709); arXiv:1706.07390v3 [hep-th] (2017).
	
	\bibitem{Bala}
	V. Balasubramanian et al., "Holographic Thermalization", Phys. Rev. D 84, 026010 (2011); arXiv:1103.2683 [hep-th].

	\bibitem{RTderiv} S. Ryu and T. Takayanagi, “Holographic derivation of entaglement entropy from the anti-de Sitter space/conformal field theory correspondence,” Phys. Rev. Lett., 96, 181602 (2006); arXiv:hep-th/0603001 (2006).
		
	\bibitem{Hubeny:2007xt}
	V.~E.~Hubeny, M.~Rangamani and T.~Takayanagi,
	``A Covariant holographic entanglement entropy proposal,''
	JHEP {\bf 0707}, 062 (2007);
	arXiv:0705.0016 [hep-th].
	
	\bibitem{Hubeny:2013hz}  V.~E.~Hubeny, M.~Rangamani and E.~Tonni,
	"Thermalization of Causal Holographic Information,"
	JHEP {\bf 1305}, 136 (2013);
	arXiv:1302.0853 [hep-th].
	
	\bibitem{Alishahiha:2014jxa} 
	M.~Alishahiha, M.~R.~M.~Mozaffar and M.~R.~Tanhayi,
	``On the Time Evolution of Holographic n-partite Information,''
	JHEP {\bf 1509}, 165 (2015);
	arXiv:1406.7677 [hep-th].
	
	\bibitem{Ben-Ami:2014gsa}
	O.~Ben-Ami, D.~Carmi and J.~Sonnenschein,
	``Holographic Entanglement Entropy of Multiple Strips,''
	JHEP {\bf 1411}, 144 (2014);
	arXiv:1409.6305 [hep-th].
	
\end{thebibliography}
\end{document}